\documentclass[useAMS,a4,usenatbib]{mn2e}
\input psfig.sty

\def \xoff {\ifmmode x_{\rm off} \else $x_{\rm off}$ \fi}
\def \rhorms {\ifmmode \rho_{\rm rms} \else $\rho_{\rm rms}$ \fi}

\def \aj  {AJ}
\def \apj  {ApJ}
\def \apjs  {ApJS}
\def \apjl  {ApJL}
\def \aap  {A\&A}
\def \prd {Phy.Rev.D}
\def \mnras {MNRAS}
\def \nat {Nature}
\def \etal {et~al.~}
\def \chisq  {\ifmmode  \chi^2   \else  $\chi^2$  \fi}  
\def \spose#1{\hbox  to 0pt{#1\hss}}  
\def \lta{\mathrel{\spose{\lower 3pt\hbox{$\sim$}}\raise  2.0pt\hbox{$<$}}}
\def \gta{\mathrel{\spose{\lower  3pt\hbox{$\sim$}}\raise 2.0pt\hbox{$>$}}}

\def \kms {\ifmmode  \,\rm km\,s^{-1} \else $\,\rm km\,s^{-1}  $ \fi }
\def \kpc {\ifmmode  {\rm kpc}  \else ${\rm  kpc}$ \fi  }  
\def \hkpc {\ifmmode  {h^{-1}\rm kpc}  \else ${h^{-1}\rm kpc}$ \fi  }  
\def \hMpc {\ifmmode  {h^{-1}\rm Mpc}  \else ${h^{-1}\rm Mpc}$ \fi  }  
\def \Msun {\ifmmode {\rm M}_{\odot} \else ${\rm M}_{\odot}$ \fi} 
\def \hMsun {\ifmmode h^{-1}\,\rm M_{\odot} \else $h^{-1}\,\rm M_{\odot}$ \fi}

\def \LCDM {\ifmmode \Lambda{\rm CDM} \else $\Lambda{\rm CDM}$ \fi}
\def \sig8 {\ifmmode \sigma_8 \else $\sigma_8$ \fi} 
\def \OmegaM {\ifmmode \Omega_{\rm m} \else $\Omega_{\rm m}$ \fi} 
\def \OmegaL {\ifmmode \Omega_{\rm \Lambda} \else $\Omega_{\rm \Lambda}$\fi} 
\def \Deltavir {\ifmmode \Delta_{\rm vir} \else $\Delta_{\rm vir}$ \fi}
\def \rhocrit {\ifmmode \rho_{\rm crit} \else $\rho_{\rm crit}$ \fi}
\def \rhou {\ifmmode \rho_{\rm u} \else $\rho_{\rm u}$ \fi}
\def \zc {\ifmmode z_{\rm c} \else $z_{\rm c}$ \fi}

\def \rhos {\ifmmode \rho_{\rm s} \else $\rho_{\rm s}$ \fi} 
\def \rmtwo {\ifmmode r_{-2} \else $r_{-2}$ \fi} 
\def \cvir {\ifmmode c_{\rm vir} \else $c_{\rm vir}$ \fi} 
\def \Rvir {\ifmmode R_{\rm vir} \else $R_{\rm vir}$ \fi}
\def \rvir {\ifmmode r_{\rm vir} \else $r_{\rm vir}$ \fi}
\def \Vvir {\ifmmode V_{\rm  vir} \else  $V_{\rm vir}$  \fi} 
\def \Mvir {\ifmmode M_{\rm  vir} \else $M_{\rm  vir}$ \fi}  
\def \Nvir {\ifmmode N_{\rm  vir} \else $N_{\rm  vir}$ \fi}  
\def \Jvir {\ifmmode J_{\rm vir} \else $J_{\rm vir}$ \fi} 
\def \Evir {\ifmmode E_{\rm vir} \else $E_{\rm vir}$ \fi} 
\def \lam {\ifmmode \lambda  \else $\lambda$ \fi} 
\def \lamp {\ifmmode \lambda^{\prime} \else $\lambda^{\prime}$  \fi} 
\def \Vmax {\ifmmode V_{\rm  max} \else  $V_{\rm max}$  \fi} 

\title[Dark halo structure in the Planck era] { Cold dark matter
  haloes in the Planck era: evolution of structural parameters for
  Einasto and NFW profiles}

\author[A.A Dutton \& A.V. Macci\`o]
{Aaron A. Dutton\thanks{dutton@mpia.de}, Andrea V. Macci\`o\\
  Max-Planck-Institut f\"ur Astronomie, K\"onigstuhl 17, 69117
  Heidelberg, Germany\\}

\begin{document} 
              
\date{Accepted 2014 April 11.  Received 2014 April 11; in original form 2014 February 27}
              
\pagerange{\pageref{firstpage}--\pageref{lastpage}}\pubyear{2014} 
 
\maketitle            

\label{firstpage}
             
\begin{abstract}  
  We present the evolution of the structure of relaxed cold dark
  matter haloes in the cosmology from the Planck satellite. Our
  simulations cover 5 decades in halo mass, from dwarf galaxies to
  galaxy clusters.  Due to the increased matter density and power
  spectrum normalization the concentration mass relation in the Planck
  cosmology has a $\sim 20\%$ higher normalization at redshift $z=0$
  compared to WMAP cosmology.  We confirm that CDM haloes are better
  described by the Einasto profile; for example, at scales near galaxy
  half-light radii CDM haloes have significantly steeper density
  profiles than implied by NFW fits.  There is a scatter of $\sim 0.2$
  dex in the Einasto shape parameter at fixed halo mass, adding
  further to the diversity of CDM halo profiles.  The evolution of the
  concentration mass relation in our simulations is not reproduced by
  any of the analytic models in the literature.  We thus provide a
  simple fitting formula that accurately describes the evolution
  between redshifts $z=5$ to $z=0$ for both NFW and Einasto fits.
  Finally, the observed concentrations and halo masses of spiral
  galaxies, groups and clusters of galaxies at low redshifts are
  in good agreement with our simulations, suggesting only mild halo
  response to galaxy formation on these scales.
\end{abstract}


\setcounter{footnote}{1}

\section{Introduction}
\label{sec:intro}

In the standard theoretical framework for structure formation in the
Universe, the mass-energy budget is dominated by a cosmological
constant and cold dark matter (CDM).  In this paradigm, initially
small density perturbations grow via gravitational instability,
forming bound structures known as dark matter haloes.

The structure of dark matter haloes are of particular interest as they
provide a non-linear scale test of the cold dark matter paradigm,
cosmological parameters, and more generally for the nature of dark
matter itself (e.g., Moore 1994; Flores \& Primack 1994; de Blok \etal
2001; Zentner \& Bullock 2002; McGaugh 2004). They also provide the
backbone for the structural properties of galaxies and galaxy scaling
relations (e.g., Mo \etal 1998; Dutton \etal 2007, 2013).

From the theoretical side there are two hurdles that need to be
overcome before an accurate prediction for the structure of CDM haloes
is possible: 1) halo structure is sensitive to cosmological parameters
(e.g., Macci\`o \etal 2008) and 2) the galaxy formation process can
cause haloes to both contract or expand (e.g., Di Cintio \etal 2014).
The subject of this paper is to constrain the effects of the
cosmological parameters on the ``baryon free'' dark halo structure, and
to quantify the evolution of the structure of CDM haloes as a
population across cosmic time.  This work continues on from our
earlier studies (Macci\`o \etal 2007, 2008; Mu{\~n}oz-Cuartas \etal
2011), as well as numerous studies in the literature (e.g., Navarro,
Frenk \& White 1996, 1997; Bullock \etal 2001; Eke \etal 2001; Zhao
\etal 2003, 2009; Duffy \etal 2008; Gao \etal 2008; Klypin \etal 2011;
Prada \etal 2012; Ludlow \etal 2013a,b).

\begin{figure}
\psfig{figure=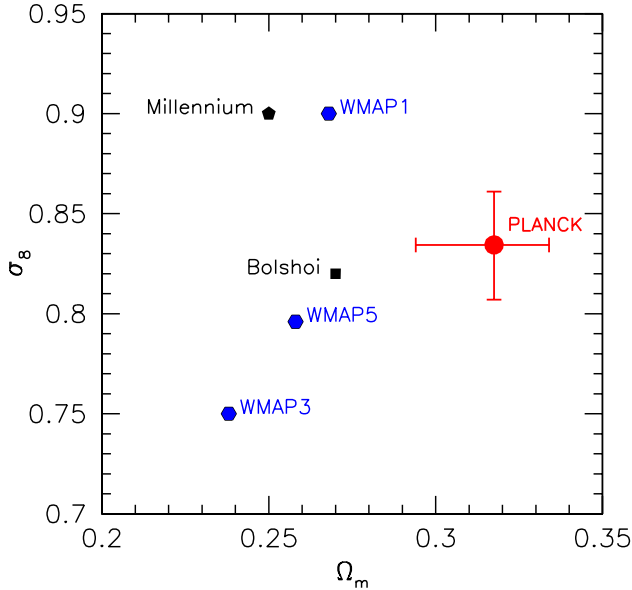,width=0.47\textwidth}
\caption{Constraints on the cosmic matter density ($\Omega_{\rm m}$)
  and the power spectrum normalization ($\sigma_8$) from the Planck
  Collaboration (2013) (red point with 68\% confidence intervals). For
  comparison, the various sets of cosmological parameters from the
  WMAP satellite are shown with blue hexagons (1st, 3rd and 5th year
  results), and the parameters used by the Millennium and Bolshoi
  N-body simulations are given by the black pentagon and square,
  respectively.}
\label{fig:omega_sigma}
\end{figure}

As shown by Macci\`o \etal (2008) relatively small changes in
cosmological parameters have a non-negligible effect on the structure
of CDM haloes. For example, the mean concentrations of CDM haloes
varied by a factor of $1.5$ between the various WMAP cosmologies
(Spergel \etal 2003, 2007). The cosmology advocated by the Planck
satellite (the Planck Collaboration 2013) has a significantly higher
matter density, $\Omega_{\rm m}$, than adopted in all previous
high-resolution simulations (See Fig.~\ref{fig:omega_sigma} and
Table~\ref{tab:cosmo_param}).  Compared to the WMAP 5th year cosmology
(Komatsu \etal 2009), the Planck cosmology also has higher
$\sigma_8$. These differences are expected to result in increased dark
halo concentrations (e.g., using the model of Bullock \etal 2001).

\begin{table*}
 \centering
   \caption{Cosmological Parameters. All cosmologies are flat, i.e.,
     $\Omega_{\Lambda}+\Omega_{\rm m}=1$. }
  \begin{tabular}{lcccccl}
\hline  
Name       &  $\Omega_{\rm m}$      & $h$             & $\sigma_8$      & $n$             & $\Omega_{\rm b}$ & Simulation Reference\\
\hline
Planck     & 0.3175           & 0.671           & 0.8344          & 0.9624           & 0.0490  &  This paper \\
WMAP5      & 0.258\phantom{1} & 0.72\phantom{1} & 0.796\phantom{1}& 0.963\phantom{1} & 0.0438 & Macci\`o \etal 2008\\
WMAP3      & 0.238\phantom{1} & 0.73\phantom{1} & 0.75\phantom{11}& 0.95\phantom{11} & 0.042\phantom{1} & Macci\`o \etal 2008\\
WMAP1      & 0.268\phantom{1} & 0.71\phantom{1} & 0.90\phantom{11}& 1.0\phantom{111} & 0.044\phantom{1} & Macci\`o \etal 2008\\
Millennium & 0.25\phantom{11} & 0.73\phantom{1} & 0.90\phantom{11}& 1.0\phantom{111} & 0.045\phantom{1} & Springel \etal 2005\\
Bolshoi    & 0.27\phantom{11} & 0.70\phantom{1} & 0.82\phantom{11}& 0.95\phantom{11} & 0.0469 & Klypin \etal 2011\\
\hline 
\end{tabular}
\label{tab:cosmo_param}
\end{table*}

Traditionally the structure of CDM haloes has been described by the
two parameter NFW profile (Navarro, Frenk, \& White 1996; 1997). This
has a divergent inner density profile of $\rho(r)\propto r^{-1}$, and
an outer profile of $\rho(r)\propto r^{-3}$. A common parametrization
uses the halo mass, $M_{200}$, and the concentration parameter,
$c\equiv r_{200}/r_{-2}$ (where $r_{200}$ is the virial radius, and
$r_{-2}$ is the scale radius).  The mass and concentration are
correlated, with a shallow slope ($c\propto M_{200}^{-0.1}$), and
small scatter ($\sigma_{\log c}\sim 0.1$), so the structure of CDM
haloes is almost scale free and described by a universal profile (NFW
97; Bullock \etal 2001).

Recent work has shown that three parameters provide a more accurate
description of spherically average CDM density profiles - especially
at small radii ($\sim 1\%$ of the virial radius). The most common
generalizations are to allow the inner logarithmic density slope,
$\gamma$, to be a free parameter (recall the NFW profile has
$\gamma=-1$), sometimes known as a generalized NFW profile (or gNFW);
or the Einasto profile (Einasto 1965), which is $d\ln\rho / d\ln r
\propto r^{\alpha}$.  A number of studies have shown that the Einasto
profile provides, in general, a better description of CDM haloes than
the NFW or gNFW profiles (e.g., Navarro \etal 2004, 2010; Merritt
\etal 2005, 2006; Stadel \etal 2009; Reed \etal 2011).  Going one step
further, using stacks of CDM haloes, Gao \etal (2008) showed that the
Einasto shape parameter, $\alpha$, depends on halo mass.

As in our earlier studies we use a large suite of cosmological N-body
simulations with different box sizes to cover the entire halo mass
range from $\sim 10^{10} \hMsun$ (haloes that host dwarf galaxies) to
$\sim 10^{15} \hMsun $ (massive clusters).  We use these simulations
to investigate the structure of CDM haloes across cosmic time.
In estimating halo concentrations we consider fits to the density
profiles using both the Einasto and NFW functions, as well as a
non-parametric approximation utilizing $V_{\rm max}/V_{200}$ following
Klypin \etal (2011).

This paper is organized as follows: in \S\ref{sec:nbody} the
simulations and the determination of the halo parameters are
presented.  In \S\ref{sec:einasto} we discuss the quality of Einasto
vs NFW fits as well as different methods for measuring halo
concentrations. In \S\ref{sec:concentration} we compare the
concentration mass relation of Planck cosmology to that from WMAP, and
discuss the effects of different cosmological parameters. In
\S\ref{sec:evolution} we present the evolution of the NFW
concentration mass relation, while in \S\ref{sec:evolution_einasto},
we present the evolution of the Einasto concentration and shape
parameters. In \S\ref{sec:observations} we compare the concentration
mass relation from our simulations with observations of spiral
galaxies and clusters of galaxies.  Finally, we summarize our results
in \S\ref{sec:summary}.

\section{N-body simulations} 
\label{sec:nbody}

Table \ref{tab:sims} lists the parameters of the simulations run in
the Planck cosmology that are used in this paper. Our simulation set
up and N-body methods follows closely that of Macci\`o \etal (2007;
2008).  We ran simulations in twelve different box sizes, which allows
us to probe halo masses covering the range $10^{10} \Msun \lta M \lta
10^{15} \hMsun$. This strategy also results in a given mass halo
appearing in several different simulations (which have different
spatial resolutions) and thus enables valuable empirical convergence
tests of halo properties. Each simulation has a unique name, which
refers to the box size in $h^{-1}{\rm Mpc}$.  In addition, in some
case we have run multiple (up to four) simulations for the same box
size in order to increase the final number of dark matter haloes and
reduce any impact of cosmic variance.

All simulations have been performed with PKDGRAV, a tree code written
by Joachim Stadel and Thomas Quinn (Stadel 2001). The code uses spline
kernel softening, for which the forces become completely Newtonian at
2 softening lengths.  Individual time steps for each particle are
chosen proportional to the square root of the softening length,
$\epsilon$, over the acceleration, $a$: $\Delta t_{\rm i} =
\eta\sqrt{\epsilon/a_{\rm i}}$. Throughout, we set $\eta = 0.2$, and
we keep the value of the softening length constant in co-moving
coordinates during each run. The physical values of $\epsilon$ at
$z=0$ are listed in Table \ref{tab:sims}.  Forces are computed using
terms up to hexadecapole order and a node-opening angle $\theta$ which
we change from $0.55$ initially to $0.7$ at $z=2$.  This allows a
higher force accuracy when the mass distribution is nearly smooth and
the relative force errors can be large. The initial conditions are
generated with the GRAFIC2 package (Bertschinger 2001).  The starting
redshifts $z_{\rm i}$ are set to the time when the standard deviation
of the smallest density fluctuations resolved within the simulation
box reaches $0.2$ (the smallest scale resolved within the initial
conditions is defined as twice the intra-particle distance).

\begin{table}
 \centering
   \caption{N-body simulation parameters for Planck cosmology runs.}
  \begin{tabular}{lccccrrrrrr}
\hline  Name &  Box  size, L  & N  &  part. mass, $m_{\rm p}$  &  force  soft., $\epsilon$ \\
& $[h^{-1}{\rm Mpc}]$ &  & $[h^{-1}{\rm M}_{\odot}]$  & $[h^{-1}{\rm kpc}]$ \\
\hline 
P-20.1   & 20   & $300^3$ & 2.611$\times10^7$ & 1.67 \\
P-20.2   & 20   & $300^3$ & 2.611$\times10^7$ & 1.67 \\
P-20.3   & 20   & $300^3$ & 2.611$\times10^7$ & 1.67 \\
P-20.4   & 20   & $300^3$ & 2.611$\times10^7$ & 1.67 \\
P-30.1   & 30   & $300^3$ & 8.811$\times10^7$ & 2.50 \\
P-30.2   & 30   & $300^3$ & 8.811$\times10^7$ & 2.50 \\
P-60     & 60   & $600^3$ & 8.811$\times10^7$ & 2.50 \\
P-45.1   & 45   & $300^3$ & 2.974$\times10^8$ & 3.75 \\
P-45.2   & 45   & $300^3$ & 2.974$\times10^8$ & 3.75 \\
P-90     & 90   & $450^3$ & 7.049$\times10^8$ & 5.00 \\
P-80     & 80   & $350^3$ & 1.052$\times10^9$ & 5.71 \\
P-130    & 130  & $450^3$ & 2.124$\times10^9$ & 7.22 \\
P-180    & 180  & $450^3$ & 5.639$\times10^9$  & 10.0 \\
P-270    & 270  & $450^3$ & 1.903$\times10^{10}$ & 15.0 \\
P-400    & 400  & $450^3$ & 6.188$\times10^{10}$ & 22.2 \\
P-600    & 600  & $600^3$ & 8.811$\times10^{10}$ & 25.0 \\
P-1000   & 1000  & $600^3$ & 4.079$\times10^{11}$ & 41.7 \\
\hline                                  
\end{tabular}
\label{tab:sims}
\end{table}

\subsection{Halo catalog} 
In all simulations, dark matter haloes are identified using a
spherical overdensity (SO) algorithm.  Candidate groups with a minimum
of $N_{\rm f}=250$ particles are selected using a FoF algorithm with
linking length $\phi = 0.2 \times d$ (the average particle
separation).  We then: (i) find the point $C$ where the gravitational
potential is minimum; (ii) determine the radius $\bar r$ of a sphere
centered on $C$, where the density contrast is $\Delta(z)$, with
respect to the {\it critical density} of the Universe, $\rhocrit(z)
=3H(z)^2/8\pi G$.  Using all particles in the corresponding sphere we
iterate the above procedure until we converge onto a stable particle
set.  For each stable particle set we obtain the virial radius,
$\rvir$, the number of particles within the virial radius, $\Nvir$,
and the virial mass, $\Mvir$.  For the Planck cosmology
$\Delta(0)\simeq 104.2$, based on the fitting function of Mainini
\etal (2003). Values of $\Delta(z)$ at other redshifts are given in
Table~\ref{tab:cm_fits}. It is also convenient to define a virial
radius, $r_{200}$, inside of which the density of the dark matter halo
is 200 times $\rhocrit$; accordingly we also define $M_{200}$ and
$N_{200}$ as the mass and the number of particles within
$r_{200}$. The advantage of this definition is that the halo mass is
independent of cosmology, i.e., for a given density profile the
corresponding $M_{200}$ is independent of cosmological parameters,
whereas $\Mvir$ depends on $\OmegaM(z)$. In the combined set of
simulations there are 92\,903 haloes with $\Nvir>500$ at redshift
$z=0$. At $z=0.5, 1, 2, 3, 4$, \& $5$ the corresponding numbers are
77\,824, 64\,100, 42\,001, 25\,694, 14\,064, and 7\,061.

\subsection{Density profile fitting}

For each SO halo in our sample we determine a set of parameters,
including the virial mass and radius, the maximum circular velocity
and the concentration parameter.  To compute the concentration of a
halo we first determine its density profile. The halo center is
defined as the location of the most bound halo particle. We compute
the density ($\rho_i$) in 50 spherical shells, spaced equally in
logarithmic radius. The minimum radius is either 1\% of $\rvir$ or 3
times the softening length (which ever is larger) and the maximum
radius is $1.2\rvir$.  Errors on the density are computed from the
Poisson noise due to the finite number of particles in each mass
shell. We fit the density profiles with two different formulae.  The
NFW profile (Navarro \etal 1997), and the Einasto profile (Einasto
1965).

The NFW profile is given by
\begin{equation}
\frac{\rho_{\rm NFW}(r)}{\rho_{-2}} = \frac{4}{(r/r_{-2})(1+r/r_{-2})^2},
\label{eq:nfw}
\end{equation}
where $r_{-2}$ is the radius where the logarithmic slope of the
density profile is $-2$ (also known as the scale radius), and
$\rho_{-2}$ is the density at the scale radius.  During the fitting
procedure we treat both $\rmtwo$ and $\rho_{-2}$ as free parameters.
Their values, and associated uncertainties, are obtained via a
$\chi^2$ minimization procedure using the Levenberg \& Marquart
method. We define the r.m.s.  of the fit as:
\begin{equation}
\rhorms = \sqrt{\frac{1}{N}\sum_i^N { (\ln \rho_i - \ln \rho_{\rm m})^2}}
\label{eq:rms}
\end{equation}
where $\rho_{\rm m}$ is the fitted NFW density distribution.
The Einasto profile (Einasto 1965) has an extra free parameter, $\alpha$, and is given by
\begin{equation}
\frac{\rho_{\rm EIN}(r)}{\rho_{-2}} = \exp\left\{-\frac{2}{\alpha}\left[ (r/r_{-2})^{\alpha} -1 \right]\right\}.
\label{eq:ein}
\end{equation}
%
\begin{figure*}
\centerline{
\psfig{figure=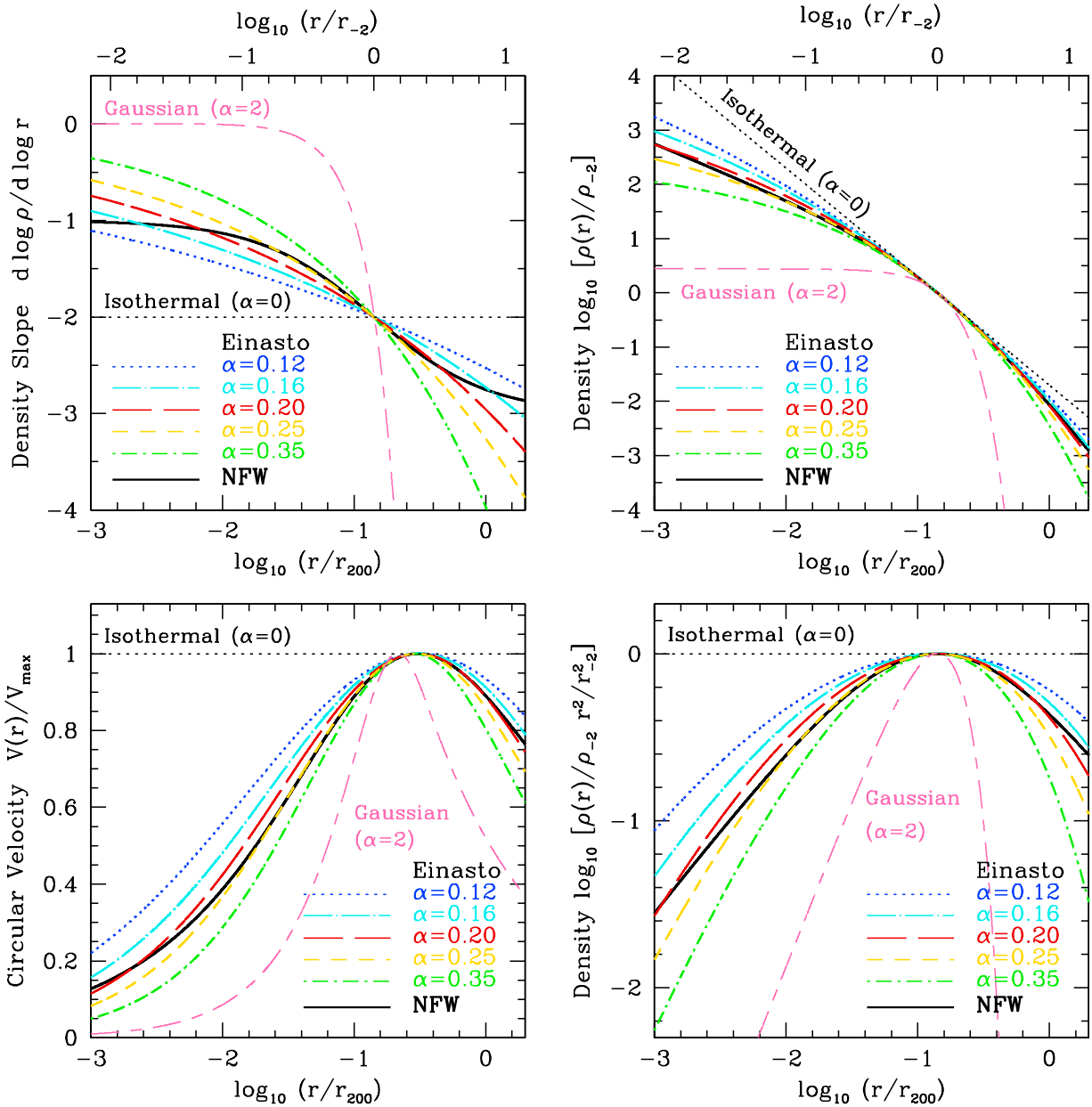,width=0.99\textwidth}
}
\caption{Einasto vs NFW profile. {\it Top Left:} density slope
  ($d\log\rho/d\log r)$), {\it Top Right:} density ($\rho(r)$), {\it
    Bottom Right:} density times radius squared ($\rho(r) r^2$), which
  displays the differences in density profiles more clearly, and is
  integrated to get the mass profile; circular velocity ($V(r)$),
  which is equivalent to cumulative mass. The NFW profile is shown
  with a solid black line. The Einasto profile is shown for values of
  the shape parameter ($\alpha$) typical of CDM haloes, as well as the
  isothermal ($\alpha=0$) and Gaussian ($\alpha=2$) cases for
  reference. In general the Einasto and NFW profiles are
  quantitatively different. However, for small radii typical of the
  optical parts of galaxies ($0.01 \lta r/r_{200}\lta 0.1$) the
  Einasto profile with $\alpha=0.25$ is very similar to the NFW
  profile, while for large radii ($0.1 \lta r/r_{200}\lta 1$) the
  Einasto profile with $\alpha=0.2$ is most similar to the NFW
  profile.}
\label{fig:einasto_vs_nfw2}
\end{figure*}

Fig.~\ref{fig:einasto_vs_nfw2} shows a comparison between the density
slopes, density, and circular velocity profiles of the NFW (solid
black lines) and Einasto (colored lines) profiles. In this example the
halo concentration $c_{200}=7.1$, which is typical for haloes of mass
$10^{13}\hMsun$ at redshift $z=0$.

The inner (as $r \rightarrow 0$) and outer (as $r \rightarrow \infty$)
slopes of the NFW profile are $-1$, and $-3$, respectively.  For the
Einasto profile the corresponding values are $0$ and $-\infty$. The
sharpness of the transition is set by the parameter $\alpha$.  In
Fig.~\ref{fig:einasto_vs_nfw2} the Einasto profile is shown for values
of the shape parameter ($\alpha$) typical of CDM haloes ($0.12 \lta
\alpha \lta 0.35$), as well as the limiting case of isothermal
($\alpha=0$), and a Gaussian ($\alpha=2$). For higher values of
$\alpha$ the Einasto density profile tends towards a uniform density
sphere of radius $r_{-2}$.

The Einasto and NFW profiles are quantitatively different at small and
large radii. However, on scales of interest for gas/stellar dynamics
and strong gravitational lensing ($0.01 \lta r/r_{200} \lta 0.1$) the
Einasto profile with $\alpha=0.25$ is very similar to the NFW profile,
and on scales of interest for weak gravitational lensing $(0.1\lta
r/r_{200}\lta 1.0)$ the Einasto profile with $\alpha=0.2$ is very
similar to the NFW profile. As shown below (see also Gao \etal 2008)
such values of $\alpha$ typically only occur for galaxy cluster mass
haloes, while galaxy mass haloes typically have $\alpha\sim 0.16$.

The concentration of the halo is defined as,
$\cvir\equiv\rvir/r_{-2}$, or $c_{200}\equiv r_{200}/r_{-2}$. Where
the virial radius (either $\rvir$ or $r_{200}$) is obtained from the
SO algorithm, and the scale radius is obtained from the density
profile fits.  We define the error on $\log c$ as
$(\sigma_{r_{-2}}/r_{-2})/\ln(10)$, where $\sigma_{r}$ is the fitting
uncertainty on $\rmtwo$.

An alternative means of estimating the concentration of the dark
matter halo utilizes the maximum circular velocity of the halo,
$V_{\rm max}$.  The circular velocity for a spherical mass
distribution is given by $V(r)=\sqrt{GM(r)/r}$.  For the NFW and
Einasto profiles there is a monotonic relation between $\Vmax/V_{200}$
and the concentration parameter, and thus $\Vmax/V_{200}$ can be used
to obtain an {\it estimate} for the concentration without explicitly
fitting the density profile. This method has been used by Klypin \etal
(2011) and Prada \etal (2012), who discovered a surprising upturn in
the concentration mass relation at high masses and redshifts.

For our simulated haloes we calculate $\Vmax$ using the same radial
grid which we use for the density profile, and we calculate the
concentration assuming the NFW relation between $V_{\rm max}/V_{200}$
and halo concentration:
\begin{equation}
\Vmax/V_{200}\simeq\sqrt{0.216 \,c_{200} / f(c_{200})},
\end{equation}
where $f(c)$ is
\begin{equation}
f(c)=\ln(1+c) -c/(1+c).
\end{equation}

An important caveat in this calculation is that the relation between
concentration and $\Vmax/V_{200}$ depends on the Einasto shape
parameter, $\alpha$. While for typical CDM halo concentrations the
Einasto profile with $\alpha=0.2$ results in very similar
concentrations as that of the NFW profile (Prada \etal 2012), CDM
haloes in general have $\alpha\ne0.2$.  It is known (e.g., Gao \etal
2008) and we also show below in \S\ref{sec:evolution_einasto}, that
the average $\alpha$ increases with halo mass and redshift.
Higher/lower values of $\alpha$ will result in higher/lower
$\Vmax/V_{200}$, and hence higher/lower concentrations when assuming
the NFW relation.  A value of $\alpha=0.2$ is typical for galaxy
cluster haloes at redshift $z=0$, and Milky Way mass haloes at
$z=3$. Thus we expect the $\Vmax/V_{200}$ method to underestimate halo
concentrations at low redshift and over-estimate them in massive
haloes at high-redshift. These expectations are realized in
\S\ref{sec:cratio}.  Additionally, $\Vmax$ is sensitive to transient
features due to unrelaxed haloes which result in an over-estimate of
the true concentration (Ludlow \etal 2012).

\subsection{Relaxed Haloes}
\label{sub:relax}
Our halo finder (and halo finders in general) does not distinguish
between relaxed and unrelaxed haloes. Unrelaxed haloes often have
poorly defined centers, which makes the determination of a radial
density profile, and hence of the concentration parameter, an
ill-defined problem.  Following Macci\`o \etal (2007; 2008) we select
relaxed haloes with the condition $\rhorms < 0.5$ and $\xoff <
0.07$. Here $\rhorms$ is the r.m.s.  of the NFW fit to the density
profile and $\xoff$ is the offset between the most bound particle and
the center of mass, in units of the virial radius $\rvir$.  This
offset is a measure of the dynamical state of the halo: relaxed haloes
in equilibrium will have a smooth, radially symmetric density
distribution, and thus an offset parameter that is virtually equal to
zero.  Unrelaxed haloes, such as those that have only recently
experienced a major merger, are likely to have a strongly asymmetric
mass distribution, and thus a relatively large $\xoff$.  At redshift
$z=0$ about 80\% of the haloes in our sample qualify as relaxed
haloes. This number decreases towards higher redshifts, reaching $\sim
50\%$ at redshift $z=5$.

\subsection{Resolution considerations}
\label{sec:resolution}
In this section we address the issue of the minimum number of
particles, $N_{\rm min}$, needed to measure reliable halo
concentrations.  We consider both an empirical determination (i.e.,
convergence of the concentration mass relation for different $N_{\rm
  min}$), and physically motivated arguments on the ability to resolve
the scale radius.

The softening length sets a minimum scale within which the enclosed
masses are not reliable (the actual resolved scale may be larger than
this).  In our simulations the softening length is set to be
$\epsilon=0.025L/N$ in co-moving coordinates, where $L$ is the box
size and $N^3$ is the number of particles. Since the particle mass is
given by $m_{\rm p}/[\hMsun]=2.775\times10^{11} (L/[\hMpc])^3 N^{-3}
\OmegaM$, the minimum halo mass by $M_{\rm min}= N_{\rm min} m_{\rm
  p}$ and the minimum viriral radius by $R_{\rm min}=(G M_{\rm
  min})^{1/3}$, we get an estimate for the maximum concentration
measurable as a function of the minimum number of particles per halo
$c_{\rm max}\equiv R_{\rm min}/\epsilon=29.0 (N_{\rm min}/1000)^{1/3}$. Since typical
halo concentrations at redshift zero are $\sim 10$, this would suggest
a few hundred particles could be sufficient to resolve the scale
radius.  Such a conclusion is backed up by our empirical convergence
studies of the concentration mass relation (see also Macci\`o \etal
2007).  For this study we adopt $N_{\rm min}=500$ when using NFW fits
which is the same as adopted by comparable studies (e.g., Klypin \etal
2011).

Since the Einasto profile has an extra free parameter there are more
degeneracies than when fitting the NFW profile. In particular, for
poorly resolved haloes there is a strong degeneracy between $\alpha$
and $c$. As such, one would expect that more particles are needed in
order to obtain reliable Einasto scale radii and shape parameters.
These expectations are realized in our empirical convergence tests of
the concentration mass relation. For Einasto fits we require several
thousand particles per halo, with more particles at lower redshifts.
The redshift dependence is expected since halo concentrations are, on
average, higher at lower redshifts, and more particles are formally
required to resolve the scale radii in haloes with higher
concentrations.  At redshift $z=0$ in the order of 10 000 particles
per halo are needed.

Power \etal (2003) give a number of convergence criteria for N-body
simulations.  For our purposes the strictest requirement is the
relaxation time (Eq. 20 in Power \etal 2003). For resolving the scale
radius, this can be approximated with $N_{\rm min}= 16 000
(c_{200}/10)^2$.  A strict application of this criteria would result in a
bias against high concentrations at masses near the resolution
limit. To get around this we select the minimum halo mass such that
84\% of the haloes at that mass satisfy the resolution requirement. 
This is given by
\begin{equation}
\label{eq:resolution}
N_{\rm min}(M_{200},z)=16 000 [1.29\bar{c}_{200}(M_{200},z)/10]^2,
\end{equation}
where $\bar{c}_{200}(M_{200})$ is obtained from our empirical fits to
the evolution of the concentration mass relation (see below). In
addition to Eq.~\ref{eq:resolution} we impose $N_{\rm min} = 3000$ for
Einasto fits. Our minimum particle numbers are thus at least as strict
as used by other authors employing Einasto fits: Gao \etal (2008)
who set $N_{\rm min}=3000$, and Ludlow \etal (2013) who set $N_{\rm
  min}=5000$.

\begin{figure}
\psfig{figure=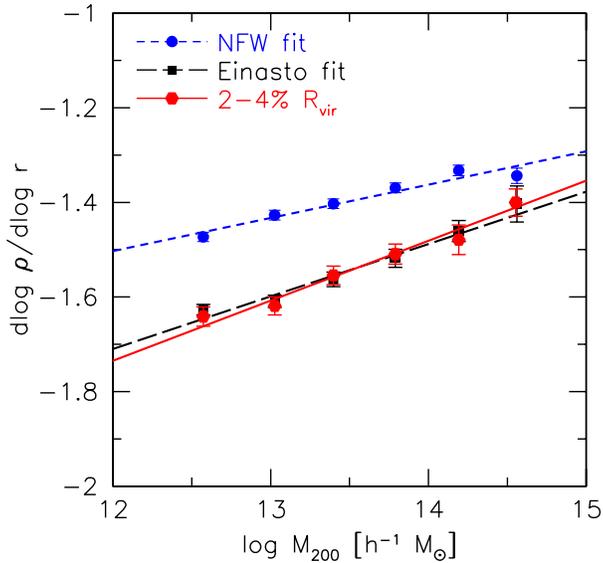,width=0.45\textwidth}
\caption{Logarithmic density slope at 3\% of the virial radius for
  relaxed haloes with more than 63,000 particles. Results from a
  direct fit from 2 to 4\% of the viriral radius are given in
  red. Results from NFW and Einasto fits to the density profiles are
  given in blue and black, respectively.  Simulated haloes are thus
  significantly steeper (at 3\% of the virial radius) than the NFW
  function, but in excellent agreement with the Einasto function.}
\label{fig:einasto_vs_nfw}
\end{figure}


\section{Einasto vs NFW}
\label{sec:einasto}

Fig.~\ref{fig:einasto_vs_nfw} shows a comparison between the
logarithmic density slopes of our simulated haloes at three percent of
the virial radius computed using three different methods. Only haloes
with more than 63 000 particles are considered here, to ensure the
profiles are fully resolved at this scale (the force softening of
these haloes is at most 0.87\% of the virial radius). For reference
the half-light radii of galaxies are of a similar scale: $\sim 3\%$ of
the virial radius for spiral galaxies, and $\sim 1.5\%$ of the viriral
radius for elliptical galaxies (Dutton \etal 2011).  The black squares
show the average slope calculated using Einasto fits to the whole
profile ($0.01 \lta r/r_{\rm vir} < 1.2$), the blue squares show the
corresponding slopes for NFW fits, while the red circles show the
slopes from a direct power-law fit to the density profile between 2
and 4\% of the virial radius.  This shows that simulated haloes have
steeper slopes (at $\sim 3\%$ of the virial radius) than the NFW fits
would suggest, but in excellent agreement with the Einasto
profile. This result agrees with previous studies at lower (Navarro
\etal 2010) and higher masses (Reed \etal 2011).  At smaller radii,
the highest resolution simulations of CDM haloes (Stadel \etal 2009;
Navarro \etal 2010) show that density profiles become monotonically
shallower inwards (consistent with the Einasto profile), with no
indication that they approach power-law behaviour. The innermost slope
measured (at $\sim 0.1\%$ of the virial radius) is shallower than the
NFW inner slope of $-1$. As shown in Fig.~\ref{fig:einasto_vs_nfw2}
the radius where the Einasto profile becomes shallower than the NFW
profile depends on $\alpha$, with smaller radii for smaller $\alpha$.
Extrapolating the Einasto profile to yet smaller radii than has
currently been resolved, predicts slopes shallower than $-0.5$ within
$\sim 10^{-5}$ virial radii, and ultimately constant density cores.

\begin{figure}
\psfig{figure=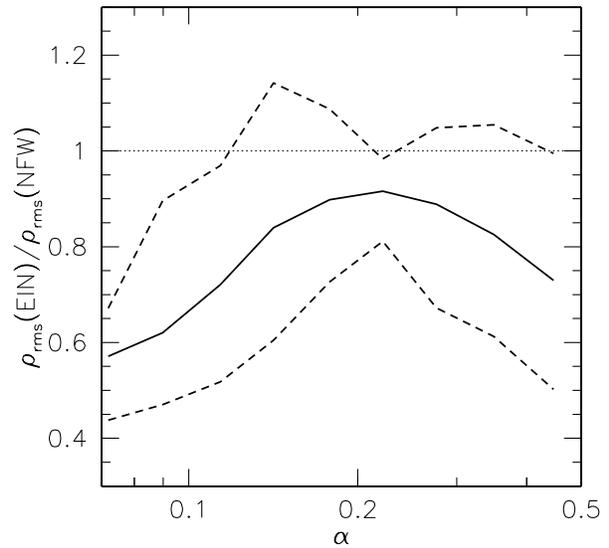,width=0.45\textwidth}
\caption{Ratio between the root mean square of deviations in $\ln\rho$
  between simulated density profiles and analytic fits using the
  Einasto and NFW profiles. The solid line shows the median, while the
  dashed lines show the 16th and 84th percentiles. The majority of
  haloes have $\rho_{\rm rms}({\rm EIN})/\rho_{\rm rms}({\rm NFW}) <
  1$ indicating the Einasto profile provides better fits for all
  values of the Einasto shape parameter, $\alpha$. Note there is a
  minimum in scatter at $\alpha\sim 0.2$, which is due to the
  similarity between the NFW and Einasto profiles over the typical
  radii resolved in our simulations.}
\label{fig:rms_ein_nfw}
\end{figure}

Another way to assess the relative quality of Einasto and NFW fits to
CDM density profiles is with the average deviation between the model
and data. Fig.~\ref{fig:rms_ein_nfw} shows a comparison of the root
mean square deviations of $\ln\rho$ for Einasto and NFW fits at
redshift $z=0$ (similar results are present at other redshifts). This
shows that for all values of the Einasto shape parameter, $\alpha$,
the majority of simulated haloes are better fit with an Einasto
profile than an NFW profile. Interestingly, as might be expected from
Fig.~\ref{fig:einasto_vs_nfw2}, the NFW fits come closest to the
Einasto fits for $\alpha\sim 0.2$, since this is where the Einasto and
NFW profiles are most similar for typical fitting regions ($0.1 \lta
r/r_{200} \lta 1$).

\begin{figure*}
\centerline{
\psfig{figure=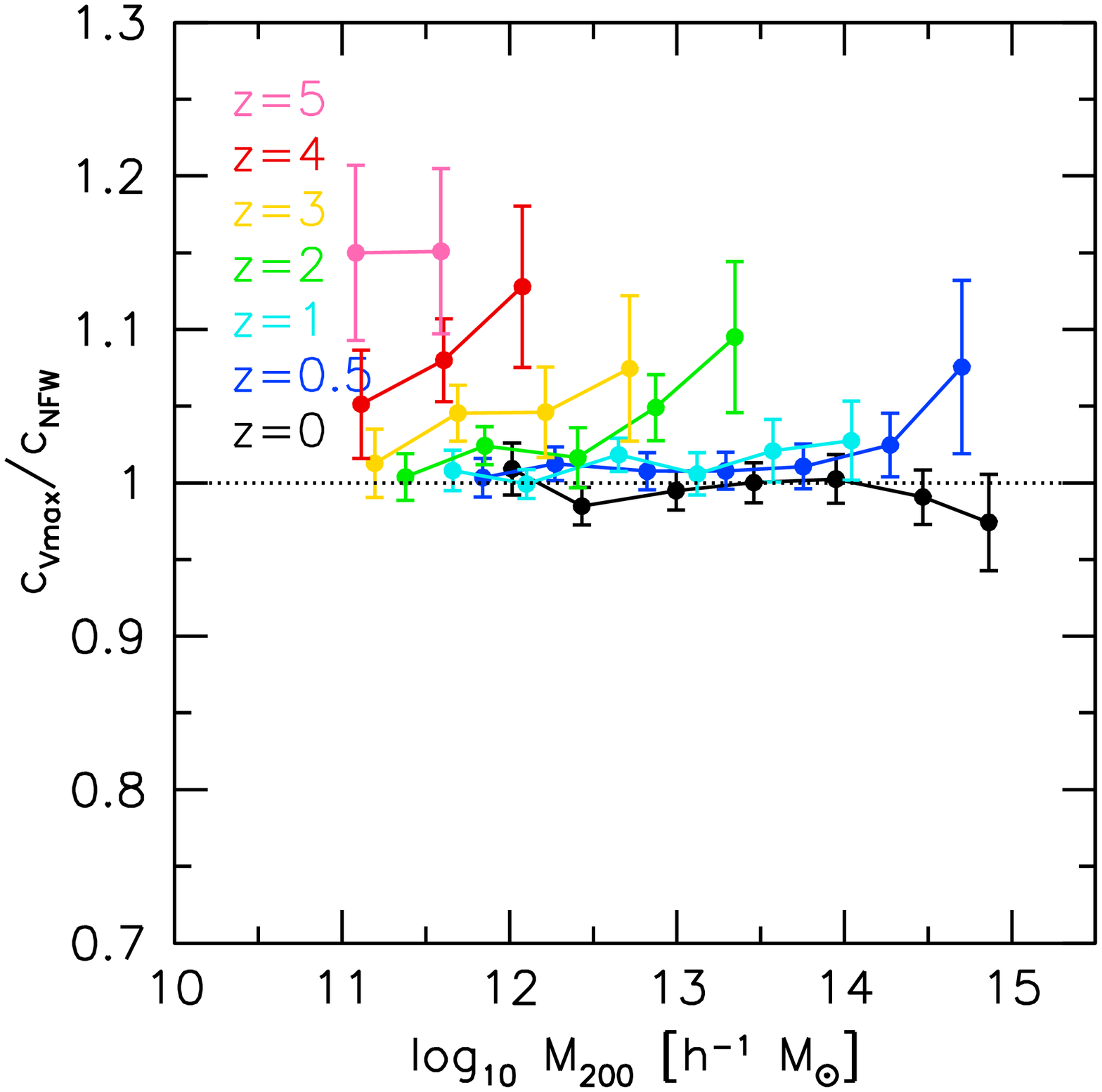,width=0.32\textwidth}
\psfig{figure=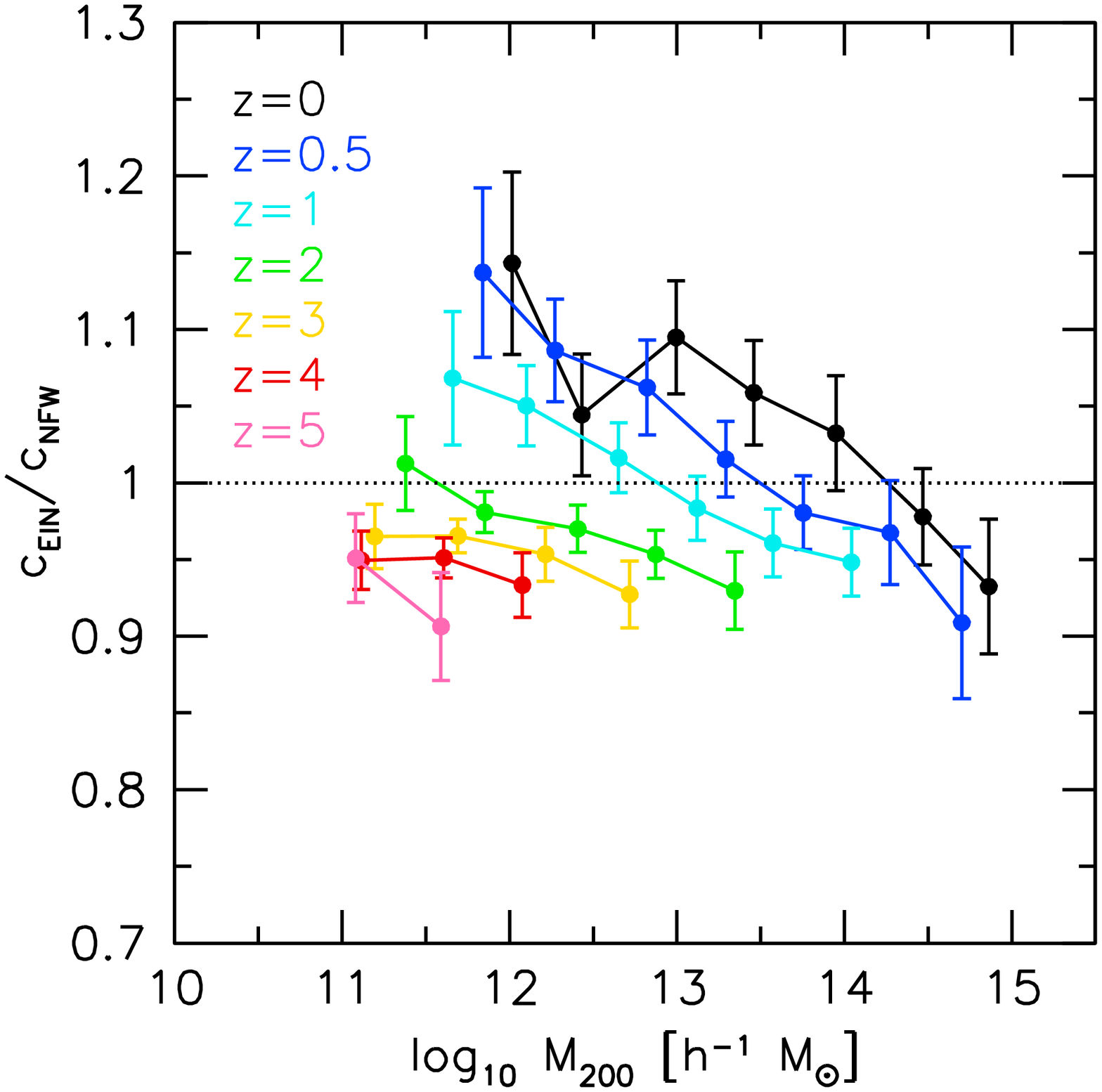,width=0.32\textwidth}
\psfig{figure=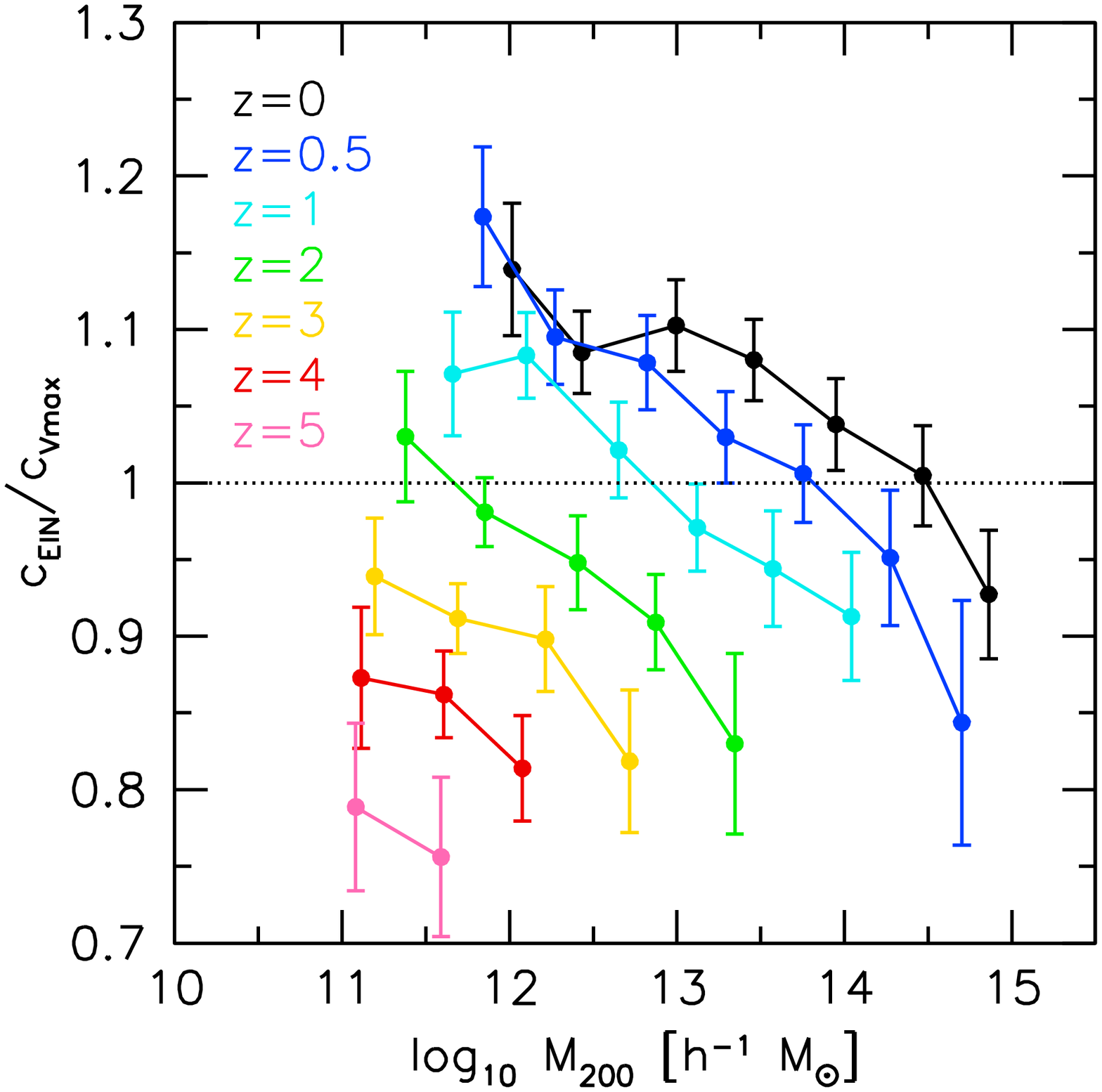,width=0.32\textwidth}
}
\caption{Dependence of halo concentrations on fitting
  procedure. Different colors represent different redshifts from $z=0$
  to $z=5$ as indicated.  Left: ratio of concentrations from
  $\Vmax/V_{200}$ and NFW fits. Middle: Ratio of concentrations from
  Einasto and NFW fits; Right: Ratio of concentrations from Einasto
  fits and $\Vmax/V_{200}$. The differences in concentrations can be
  as large as $25\%$}
\label{fig:cratio}
\end{figure*}

\subsection{Comparison between concentration estimates}
\label{sec:cratio}
In this paper we consider three different measurements of the halo
concentration: NFW fits to the density profile ($c_{\rm NFW}$),
Einasto fits to the density profile ($c_{\rm EIN}$), and the
$\Vmax/V_{200}$ ratio assuming an NFW profile ($c_{V_{\rm max}}$).  In
Fig.~\ref{fig:cratio} we show the ratio between two concentration
estimates, for the three combinations: $c_{\Vmax}/c_{\rm NFW}$ (left);
$c_{\rm EIN}/c_{\rm NFW}$ (middle); and $c_{\Vmax}/c_{\rm EIN}$
(right).  The symbols show the median while the error bars show the
statistical uncertainty on the median. Since the halo radii are the
same for each method the concentration ratio is equivalent to the
ratio of the scale radii. In order to compare all concentrations
equally we use the stricter Einasto fit particle number requirements.

\subsubsection{\Vmax vs NFW}
At redshift zero the NFW and $\Vmax$ methods give very similar
results, the differences are consistent with statistical
uncertainties. However, at higher redshifts and especially higher
masses, the $\Vmax$ method gives systematically higher concentrations,
by up to 15\%. These results are consistent with the tests from Prada
\etal (2012) (see their Figure 4), which show the $\Vmax$ method gives
higher concentrations than NFW fits, especially for low concentration
haloes.  Such a bias is expected since any perturbations from a NFW
profile (either from real features or statistical uncertainties) will
preferentially scatter $\Vmax$ up.  Additionally, for lower
concentrations the velocity ratio approaches unity ($\Vmax/V_{200}=1$
for $c_{200}\simeq 2.2$), and thus errors in the velocity ratio
propagate into larger uncertainties in concentrations.  Since halo
concentrations are lower at higher masses and redshifts, any
systematic effects are amplified.

\begin{figure}
\psfig{figure=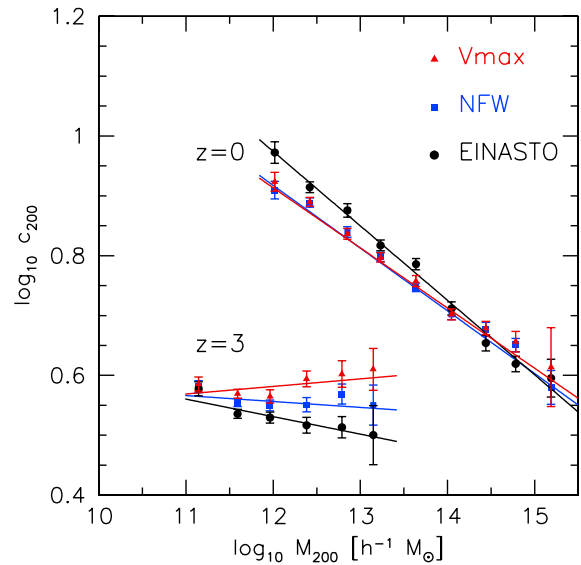,width=0.45\textwidth}
\caption{Dependence of the halo concentration mass relation on fitting
  procedure. The colors and point types indicate the different
  methods: Einasto fits (black circles); NFW fits (blue squares);
  $\Vmax/V_{200}$ (red triangles).  Notice that the Einasto fits give
  a steeper slope to the concentration mass relation at redshift
  $z=0$, while the $\Vmax$ method gives a positive slope to the
  concentration mass relation at redshift $z=3$.}
\label{fig:call}
\end{figure}

\subsubsection{NFW vs Einasto}
The deviations between NFW and Einasto concentrations are within $\sim
15\%$ at all redshifts. There is a clear trend between the
concentration ratio and halo mass, which translates into steeper
slopes for the Einasto concentration mass relation, as well as
stronger evolution (Fig.~\ref{fig:call}, and \S's~\ref{sec:evolution}
\& \ref{sec:evolution_einasto}).  These trends are qualitatively
consistent with the mass and redshift dependence of the Einasto shape
parameter.

\subsubsection{\Vmax vs Einasto}
The \Vmax and Einasto concentrations show the largest differences, up
to 25\%. As well as the biases towards higher concentrations from the
$\Vmax$ method discussed above, another cause of differences is the
fact that for an Einasto profile the conversion between
$\Vmax/V_{200}$ and $c$ depends not only on $c$ (as is the case for an
NFW profile) but also on the Einasto shape parameter, $\alpha$, with
higher $\alpha$ giving higher $\Vmax/V_{200}$.  As is known (Gao \etal
2008) higher sigma haloes have higher $\alpha$, which leads to an
over-estimate of the halo concentration for high-sigma haloes and
an underestimate for low-sigma haloes.

\begin{figure*}
\centerline{
\psfig{figure=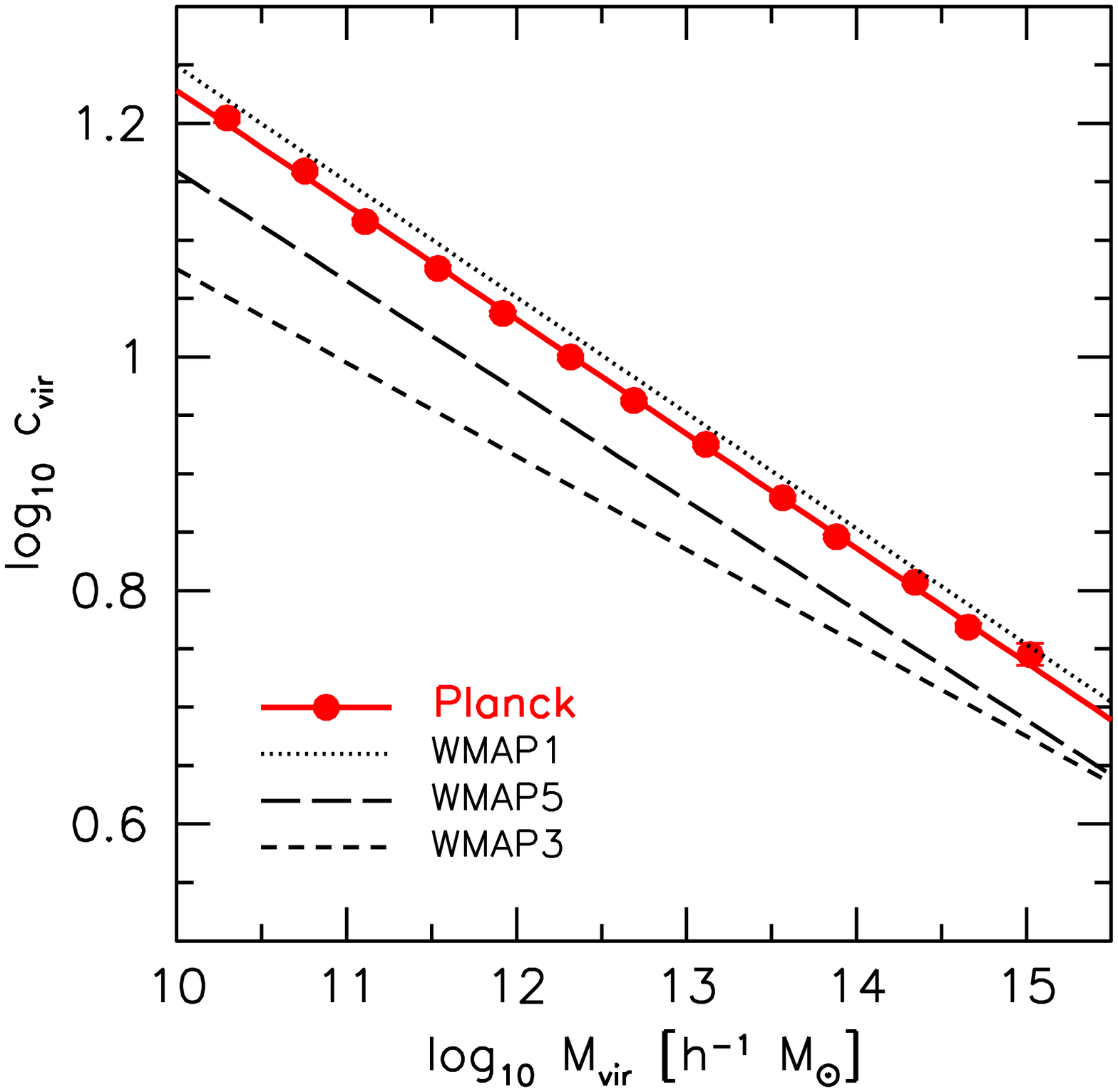,width=0.45\textwidth}
\psfig{figure=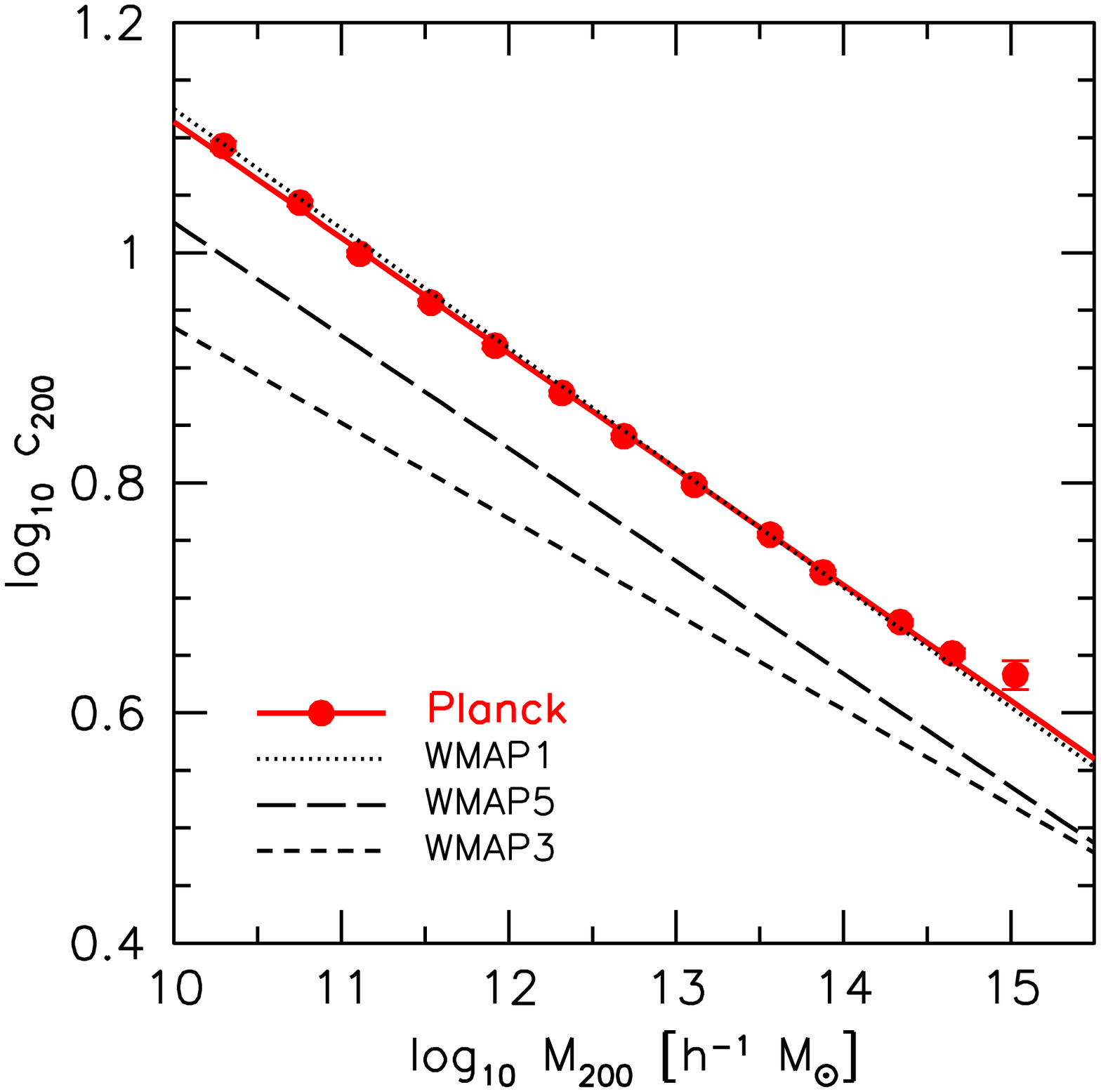,width=0.45\textwidth}
}
\caption{Concentration mass relations for the Planck cosmology (red
  circles), for the ``virial'' (left) and $200\rhocrit$ (right)
  definitions of halo masses and sizes. Halo concentrations are from
  NFW fits to the density profile. Power-law fits are shown with red
  lines. For comparison, fits to the relations for WMAP cosmologies
  from Macci\`o \etal (2008) are shown with black lines: WMAP1
  (dotted), WMAP3 (short dashed), and WMAP5 (long-dashed). The
  concentration mass relation for the Planck cosmology is remarkably
  similar to that from WMAP1, in spite of the different cosmological
  parameters (see Table~\ref{tab:cosmo_param}). Note that with the
  ``virial'' halo definition the over density of the halo with respect
  to the critical density depends on $\OmegaM$, and thus varies
  between cosmologies.}
\label{fig:cm_planck_vs_wmap}
\end{figure*}

Klypin \etal (2011) and Prada \etal (2012) used the velocity ratio
method and found an upturn in the concentration mass relation for the
highest mass haloes at all redshifts. As shown in Fig.~\ref{fig:call},
we recover an upturn at $z=3$ when using $\Vmax/V_{200}$
concentrations, but the relation is almost flat when using NFW
concentrations, and has a negative slope when using Einasto
concentrations. We suggest that the $\Vmax/V_{200}$ method could be
improved by assuming the Einasto profile with a mass dependent
$\alpha$ as specified below in \S\ref{sec:evolution_einasto}.


\section{The concentration - mass relation at redshift zero}
\label{sec:concentration}

We start our analysis with the concentration mass relation at redshift
zero. Here we use NFW fits to allow for better comparison with
previous studies. In \S\ref{sec:evolution_einasto} we give the results
for Einasto fits, which we recommend since they provide a more
accurate description of CDM haloes.

\subsection{Planck vs WMAP}

The concentration mass relations for relaxed haloes ($\xoff < 0.07$,
$\rhorms < 0.5$ ) with more than 500 particles at redshift $z=0$ in
the Planck cosmology are shown in Fig.~\ref{fig:cm_planck_vs_wmap}.
The left panel shows the relation for our fiducial ``virial'' halo
definition, while the right panel shows the relation for the
$200\rhocrit$ definition. 
The concentration mass relations are well fitted by a power-law:
\begin{equation}
\label{eq:cm}
\log_{10} c = a + b \log_{10}(M/[10^{12}h^{-1}\Msun]),
\end{equation}
so that $a$ is the zero point, and $b$ is the slope.  The best fit
relations are shown by the solid red lines in
Fig.~\ref{fig:cm_planck_vs_wmap} and are given by
\begin{equation}
\log_{10} c_{200} = 0.905 -0.101 \log_{10}(M_{200}/[10^{12}h^{-1}\Msun]),
\end{equation}
and
\begin{equation}
\log_{10} c_{\rm vir} = 1.025 -0.097 \log_{10}(\Mvir/[10^{12}h^{-1}\Msun]).
\end{equation}

The dotted, dashed and long-dashed lines show the corresponding
relations for the WMAP 1st, 3rd and 5th year cosmologies (as presented
in Macci\`o \etal 2008). For $c_{200}$ the zero points were 0.917,
0.769, and 0.830, for the WMAP 1st, 3rd, and 5th year cosmologies,
respectively.  Note that for the virial halo definition (i.e.,
$\Mvir$) the virial overdensity is a function of the matter density,
and thus varies between the different cosmologies. However, the
$200\rhocrit$ (i.e., $M_{200}$) definition is independent of
cosmology, and thus the real cosmology induced structural differences
are straight forward to see.

Our main result is that the concentrations in the Planck cosmology are
$\sim 20\%$ higher than in the WMAP 5th year cosmology.  Quite
remarkably, the concentration mass relation in the Planck cosmology is
very similar to that of the WMAP 1st year cosmology, despite the
significantly different cosmological parameters (see
Table~\ref{tab:cosmo_param}).  This similarity has been shown
previously, using an analytic model, by Ludlow \etal (2013). In
\S\ref{sec:bullock} below we show how the changes in the various
cosmological parameters effect the concentration mass relation, and
cancel each other out.

Finally, the distribution of concentrations around the mean
concentration - mass relation is well described by a log-normal with
an intrinsic scatter of 0.11 dex, consistent with results from
different cosmologies (Macci\`o \etal 2008).

\begin{figure}
\psfig{figure=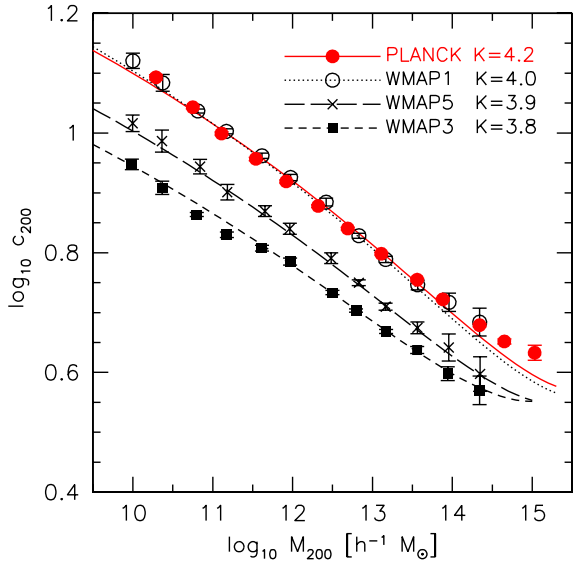,width=0.45\textwidth}
\caption{Comparison between concentration mass relations from
  simulations (symbols) with predictions from the analytic model of
  Bullock \etal (2001) as modified by Macci\`o \etal (2008). Note that
  the value of $K$ needs to vary slightly (by up to 10\%) to provide
  the best match to the simulations.}
\label{fig:bullock}
\end{figure}

\begin{figure}
\psfig{figure=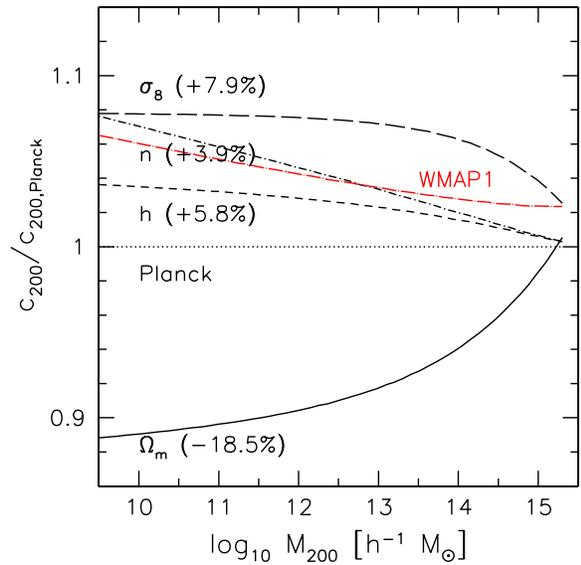,width=0.45\textwidth}
\caption{Effect of cosmological parameters on halo concentrations,
  relative to those from the Planck cosmology. This calculation was
  made using the model from Bullock \etal (2001). The various black
  lines show the effect of changing a single cosmological parameter
  from the Planck value to the WMAP1 value.  The red line uses the
  WMAP1 parameters. This shows that higher $\sigma_8$, $n$, and $h$,
  are offset by a lower $\OmegaM$.}
\label{fig:ratio}
\end{figure}

\subsection{Comparison to analytic models}
\label{sec:bullock}
Fig.~\ref{fig:bullock} shows the concentration mass relations for our
simulations (symbols) compared to the analytic model of Bullock \etal
(2001) as modified by Macci\`o \etal (2008).  As found by Macci\`o et
al. (2008), for this model to reproduce the simulations exactly, the
parameter $K$ needs to vary slightly (by a few percent) for different
cosmologies. We find that $K\simeq 4.2$ provides the best match for
the Planck cosmology. With the larger volumes for our simulations in
the Planck cosmology we probe a factor of $\sim 6$ higher in halo
mass, at these scales ($M_{200}\gta 3\times 10^{14}\hMsun$) we see a
departure from the Bullock model.

Given that the parameters are so different between the Planck and
WMAP1 cosmologies: $\OmegaM=0.3175$ vs $0.268$ ($+18.5$\%); $h=0.671$ vs
$0.71$ ($-5.5$\%); $\sigma_8=0.8344$ vs $0.90$ ($-7.3$\%); $n=0.9624$ vs
$1.0$ ($-3.8$\%), we ask why are the Planck and WMAP1 concentration mass
relations so similar? Fig.~\ref{fig:ratio} shows the effect on halo
concentrations (relative to the Planck cosmology) by changing various
cosmological parameters. Reducing the matter density ($\OmegaM$)
reduces the concentrations, while increasing the Hubble parameter
($h$), power spectrum normalization ($\sigma_8$), and power spectrum
slope ($n$) increases the concentrations. The changes roughly cancel
out, leaving an expected net increase of $2-7\%$ (depending on halo
mass).  The differences in the relations from our simulations are
slightly smaller, but of a similar amplitude.

\begin{table}
 \centering
 \caption{Fit parameters for the concentration mass relation: $\log_{10} c = a + b \log_{10}(M/[10^{12}h^{-1}\Msun])$.}
  \begin{tabular}{ccccc}
\hline 
\hline 
Profile & $\Delta$ & Redshift & zero point ($a$) & slope ($b$)     \\
\hline
NFW     & 200     & 0.0       & 0.905 $\pm$0.001 & -0.101 $\pm$0.001\\       
NFW     & 200     & 0.5       & 0.814 $\pm$0.001 & -0.086 $\pm$0.001\\       
NFW     & 200     & 1.0       & 0.728 $\pm$0.001 & -0.073 $\pm$0.001\\       
NFW     & 200     & 2.0       & 0.612 $\pm$0.001 & -0.050 $\pm$0.001\\       
NFW     & 200     & 3.0       & 0.557 $\pm$0.003 & -0.021 $\pm$0.002\\       
NFW     & 200     & 4.0       & 0.528 $\pm$0.004 &  0.000 $\pm$0.003\\       
NFW     & 200     & 5.0       & 0.539 $\pm$0.006 &  0.027 $\pm$0.005\\ 
\hline      
NFW     & 104.2   & 0.0       & 1.025 $\pm$0.001 & -0.097 $\pm$0.001\\       
NFW     & 138.4   & 0.5       & 0.884 $\pm$0.001 & -0.085 $\pm$0.001\\       
NFW     & 156.9   & 1.0       & 0.775 $\pm$0.001 & -0.073 $\pm$0.001\\       
NFW     & 170.6   & 2.0       & 0.643 $\pm$0.001 & -0.051 $\pm$0.001\\       
NFW     & 174.5   & 3.0       & 0.581 $\pm$0.002 & -0.023 $\pm$0.002\\       
NFW     & 176.0   & 4.0       & 0.550 $\pm$0.004 & -0.002 $\pm$0.003\\       
NFW     & 177.3   & 5.0       & 0.559 $\pm$0.006 &  0.024 $\pm$0.005\\
\hline       
Einasto & 200     & 0.0       & 0.978 $\pm$0.006 & -0.125 $\pm$0.004\\
Einasto & 200     & 0.5       & 0.884 $\pm$0.005 & -0.117 $\pm$0.004\\
Einasto & 200     & 1.0       & 0.775 $\pm$0.004 & -0.100 $\pm$0.004\\
Einasto & 200     & 2.0       & 0.613 $\pm$0.004 & -0.073 $\pm$0.006\\
Einasto & 200     & 3.0       & 0.533 $\pm$0.005 & -0.027 $\pm$0.009\\
Einasto & 200     & 4.0       & 0.481 $\pm$0.009 & -0.020 $\pm$0.014\\
Einasto & 200     & 5.0       & 0.478 $\pm$0.022 &  0.013 $\pm$0.032\\
\hline 
\hline
\end{tabular}
\label{tab:cm_fits}
\end{table}

\begin{figure*}
\centerline{
\psfig{figure=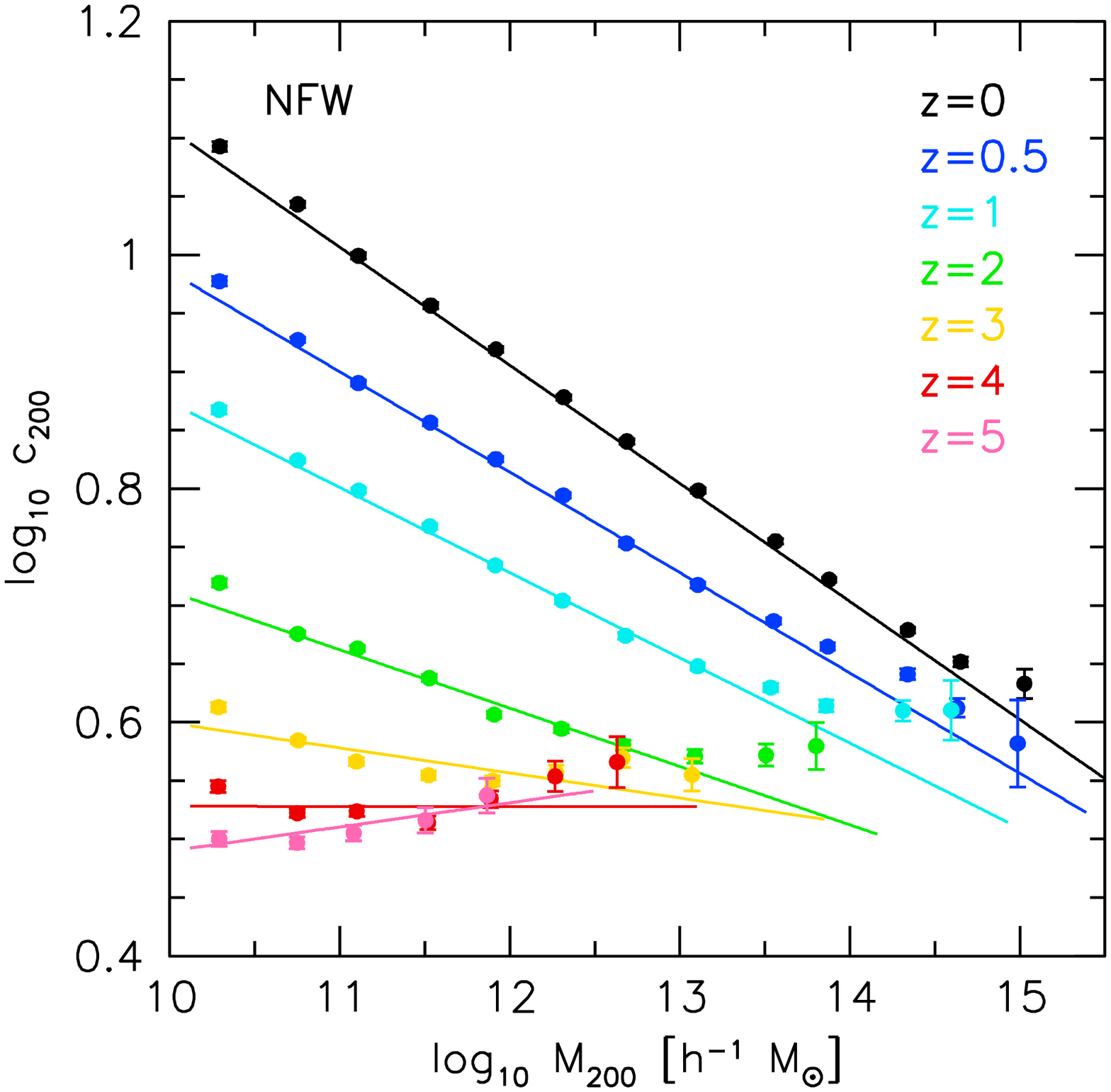,width=0.45\textwidth}
\psfig{figure=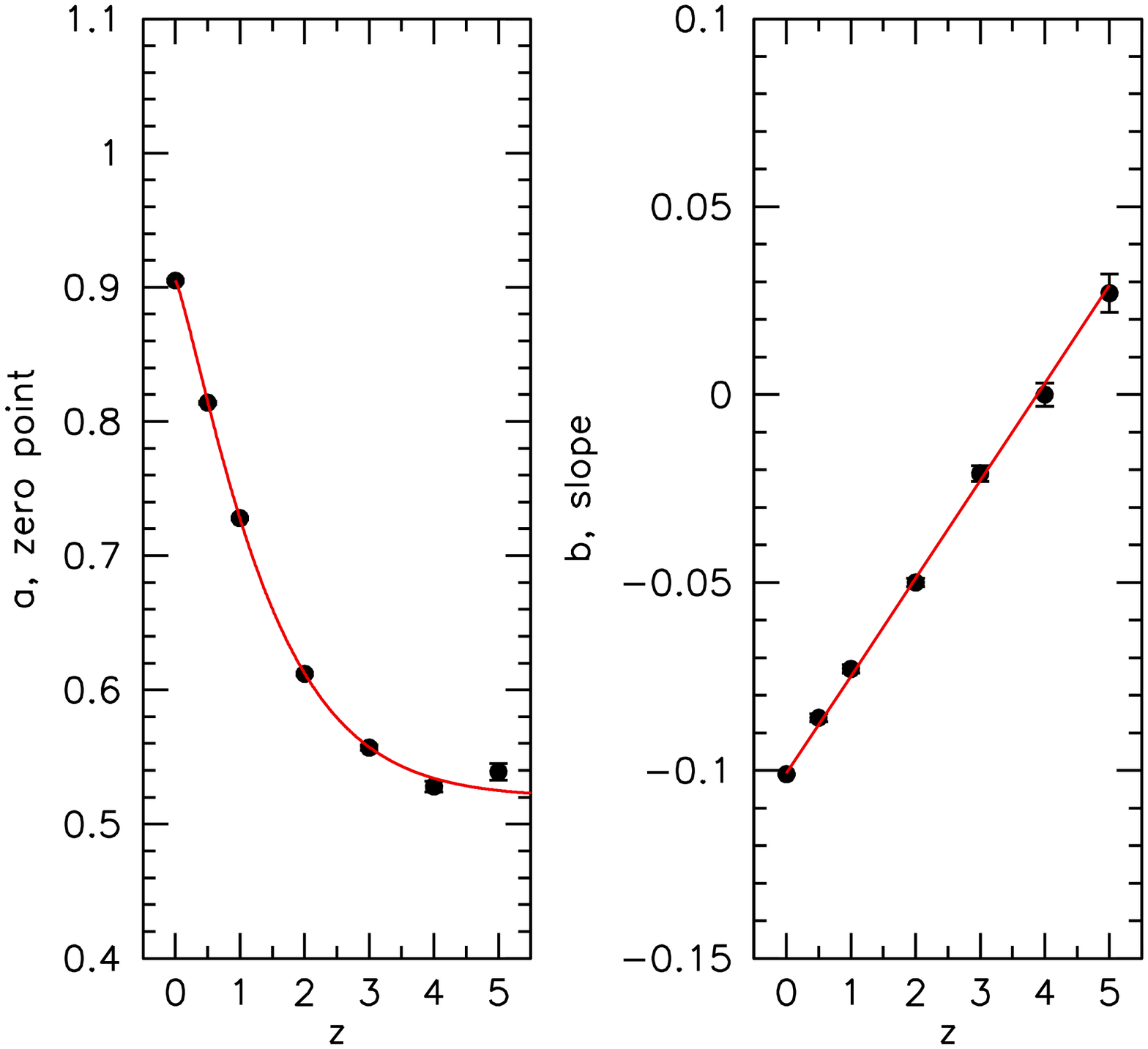,width=0.45\textwidth}
}
\centerline{
\psfig{figure=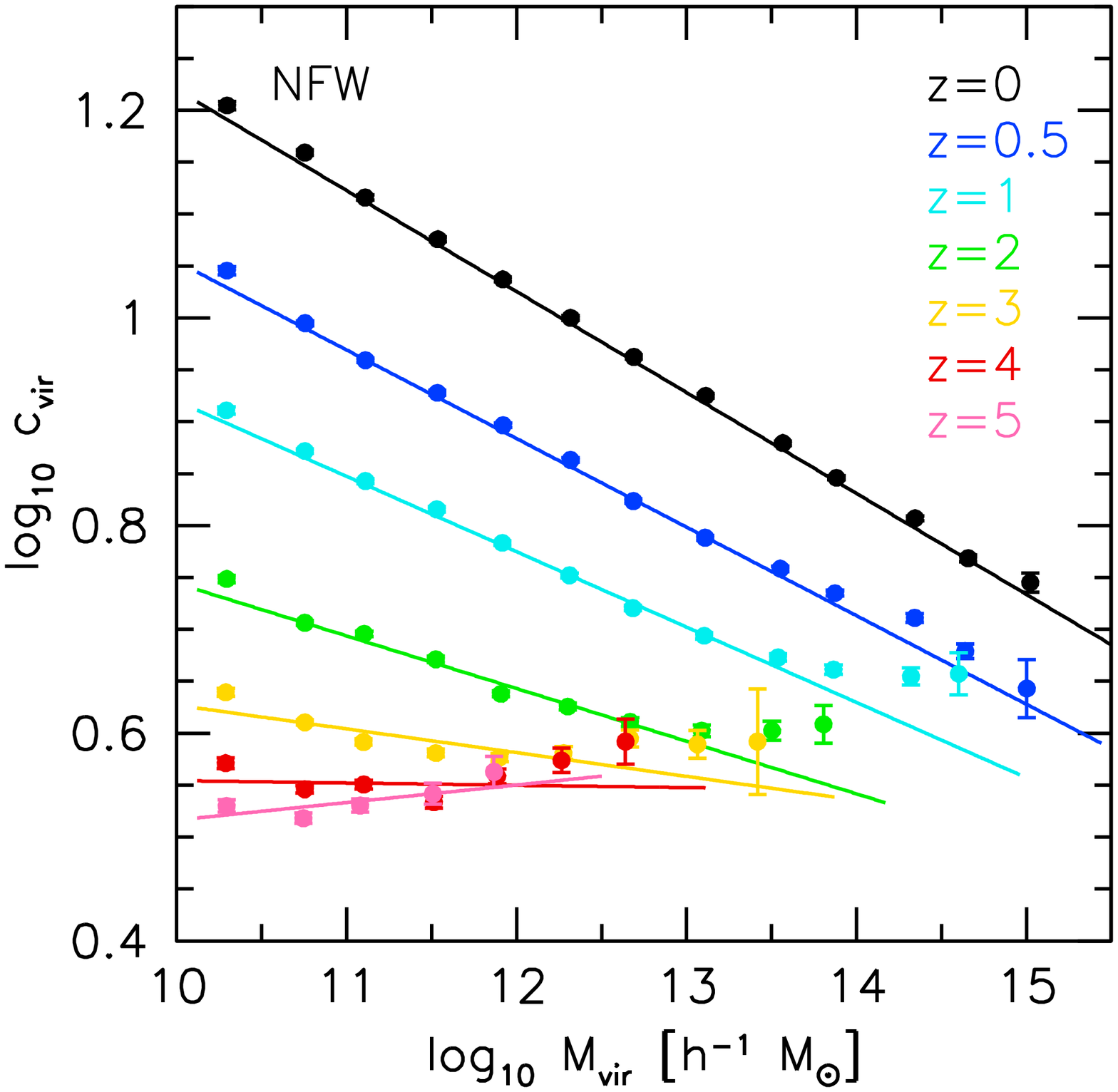,width=0.45\textwidth}
\psfig{figure=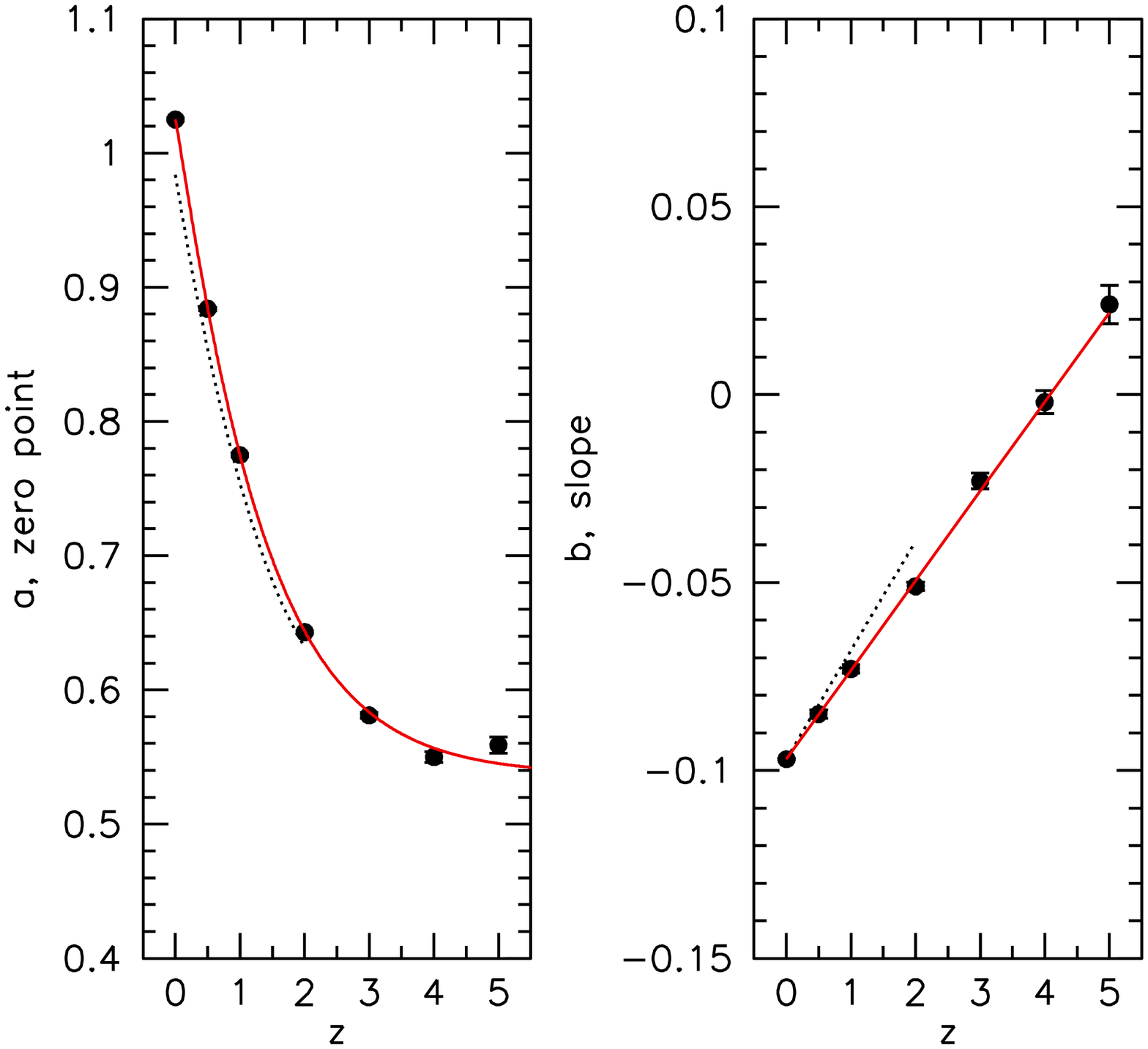,width=0.45\textwidth}
}
\caption{Evolution of the concentration mass relation using NFW fits.
  Upper panels show results for $c_{200}=r_{200}/r_{-2}$, while lower
  panels show results for $\cvir=\rvir/r_{-2}$.  The left panels show
  the median concentration in bins of halo mass from redshifts $z=0$
  to $z=5$ (with colors as indicated). The solid lines are power-law
  fits of the form $\log_{10}c = a + b\log_{10}(M/10^{12}\hMsun)$. The
  parameters of these fits are given as black points in the right
  panels, with the error bars showing $1\sigma$ uncertainties.  The
  evolution of the slope is linear with redshift, while the evolution
  of the zero point is approximately exponential with redshift. The
  red solid lines give fitting formula for the evolution (see
  Eqs.~\ref{eq:c200_slope}--\ref{eq:cvir_zero}). For reference, the
  dotted lines give the evolution (from $z=0$ to $z=2$) in the WMAP5
  cosmology from Mu{\~n}oz-Cuartas \etal (2011).}
\label{fig:cmz_nfw}
\end{figure*}

Since the model of Bullock \etal (2001) provides a good description of
the relative changes in halo concentrations for different cosmological
parameters we can use the model to estimate the uncertainty on
concentrations from uncertainties in the cosmological parameters.  We
use posterior chains for the base/planck\_lowl model, publicly
available from the Planck Collaboration. This model assumes a flat
cosmology. We pass the values of $\OmegaM$, $H_0$, $\sigma_8$, $n$,
and $\Omega_{\rm b}h^2$ through the Bullock \etal (2001) model. The
resulting scatter in $c_{200}$ at a halo mass of
$M_{200}=10^{12}\hMsun$ is just 0.015 dex, or 3.5\%. As we saw above
in \S\ref{sec:cratio} the systematic differences in measuring halo
concentrations are often larger than this. Furthermore, the intrinsic
scatter in halo concentrations is almost an order of magnitude
larger. Thus uncertainties on cosmological parameters are no longer a
significant source of systematic uncertainty in the prediction of
concentrations of baryon free cold dark matter haloes.

\begin{figure*}
\centerline{
\psfig{figure=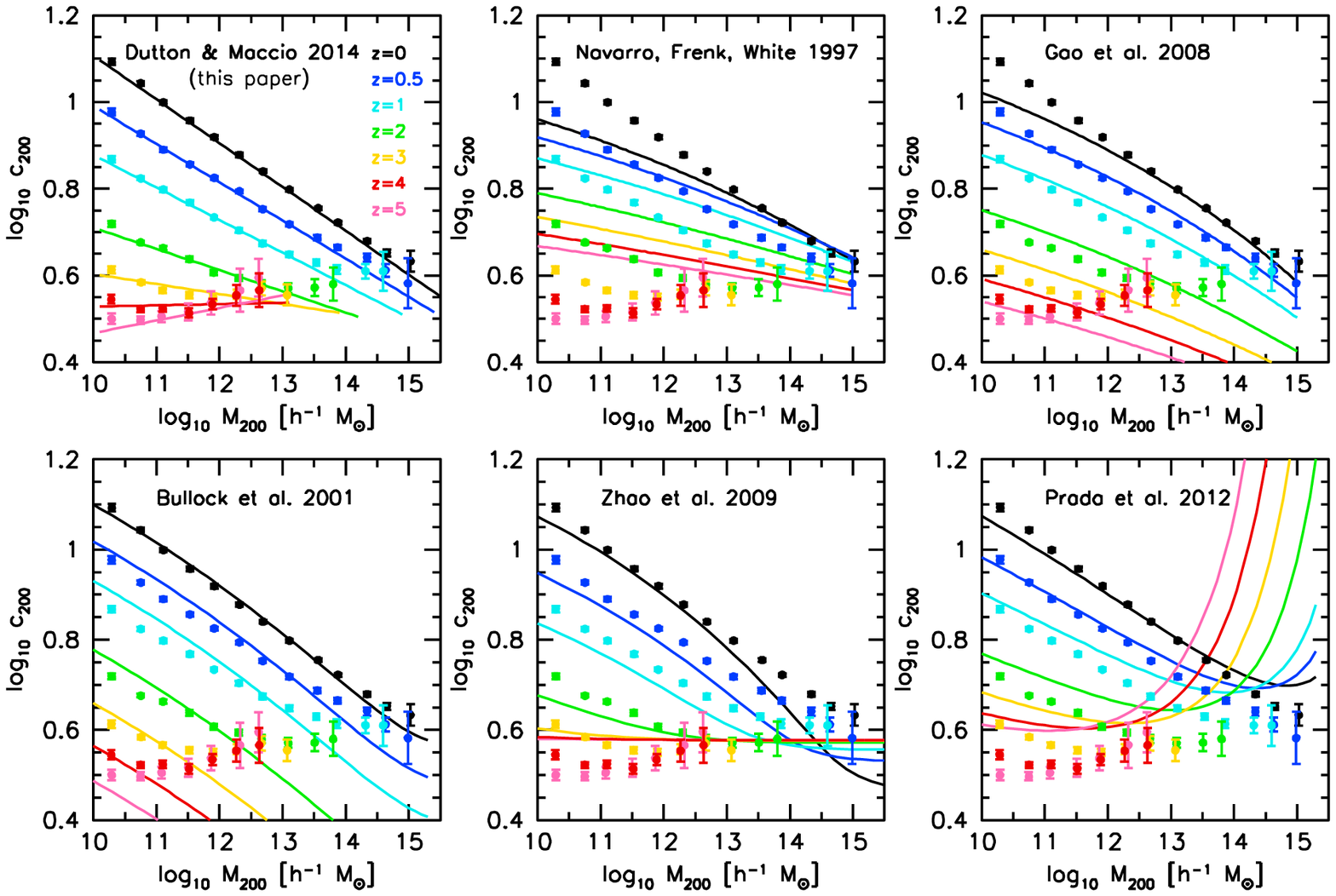,width=0.99\textwidth}
}
\caption{Comparison between concentration mass relations from
  simulations (using NFW fits) at redshifts $z=0$ to $z=5$ (points and
  errors on the median) with our fitting formula (upper left) together
  with predictions from the analytic models of Navarro, Frenk, \&
  white (1997); Bullock \etal (2001), as modified by Macci\`o \etal
  (2008); Gao \etal (2008); Zhao \etal (2009); and Prada \etal
    (2012). }
\label{fig:nfw4}
\end{figure*}

\section{Evolution of NFW concentrations}
\label{sec:evolution}

The evolution of the concentration mass relation for NFW fits is shown
in Fig.~\ref{fig:cmz_nfw}. Upper panels show $c_{200}$ vs $M_{200}$,
while lower panels show $\cvir$ vs $\Mvir$.  The left hand panels show
the median concentration mass relations from redshifts $z=5$ to
$z=0$. At all redshifts the relations are well fitted by a power-law,
except at the very highest masses, where there is some evidence for a
departure from a power-law -- specifically a flattening and sometimes
an upturn in the median concentration (see also Klypin \etal 2011;
Prada \etal 2012). However, as we discuss above in \S\ref{sec:cratio},
the size of this departure is dependent on the method used to measure
halo concentrations: e.g., NFW fits, Einasto fits, $V_{\rm
  max}/V_{200}$. The $\Vmax$ method results in the largest up-turn,
while Einasto fits show no evidence for an up-turn. Additionally, as
shown by Ludlow \etal (2012) the use of $V_{\rm max}/V_{200}$ as a
concentration estimator is especially sensitive to transient features
due to unrelaxed haloes.  We thus caution against over-interpretation
of this feature.

The panels on the right show the evolution of the slope and
normalization of the concentration mass relation fitted using a
power-law as parametrized by Eq.~\ref{eq:cm}.  The slope changes
linearly with redshift, while the zero point approaches a constant
value above $z\sim 3$. The data for the slopes and zero points are
given in Table~\ref{tab:cm_fits}.
Fits to the evolution of the slope and zero points are shown with
solid red lines. For the $c_{200}$ vs $M_{200}$ relation these are
given by
\begin{equation}
\label{eq:c200_slope}
b = -0.101 +0.026z	
\end{equation}
\begin{equation}
\label{eq:c200_zero}
a = 0.520 +(0.905-0.520) \exp(-0.617 z^{1.21}),
\end{equation}
and for the $\cvir$ vs $\Mvir$ relation by
\begin{equation}
\label{eq:cvir_slope}
b = -0.097 +0.024z	
\end{equation}
\begin{equation}
\label{eq:cvir_zero}
a = 0.537 +(1.025-0.537) \exp(-0.718 z^{1.08}).
\end{equation}
For comparison purposes the fitting formula (from $z=2$ to $z=0$) for
the $\cvir-\Mvir$ relation in the WMAP5 cosmology form
Mu{\~n}oz-Cuartas \etal (2011) is given by dotted lines. The form of
the evolution in slope and zero point is qualitatively the same,
although the details differ.

\subsection{Comparison with analytic models}
A comparison between the results of our N-body simulations and
predictions of several analytic models are shown in
Fig.~\ref{fig:nfw4}. The upper left panel shows our fitting formula
(Eqs.~\ref{eq:c200_slope} \& \ref{eq:c200_zero}).
%
The upper middle panel shows the Navarro, Frenk, \& White (1997) model
using the original parameters $F=0.5$, $f=0.01$, $C=3000$. The success
of this model is in reproducing the concentrations of massive
$M_{200}\gta 10^{13}\hMsun$ haloes at redshift $z=0$. At all other
masses and redshifts, however, this model does a poor job at
reproducing our simulation results. Specifically it predicts much
weaker evolution in the slope and zero point than our simulations (see
also Bullock \etal 2001).

\begin{figure*}
\centerline{
\psfig{figure=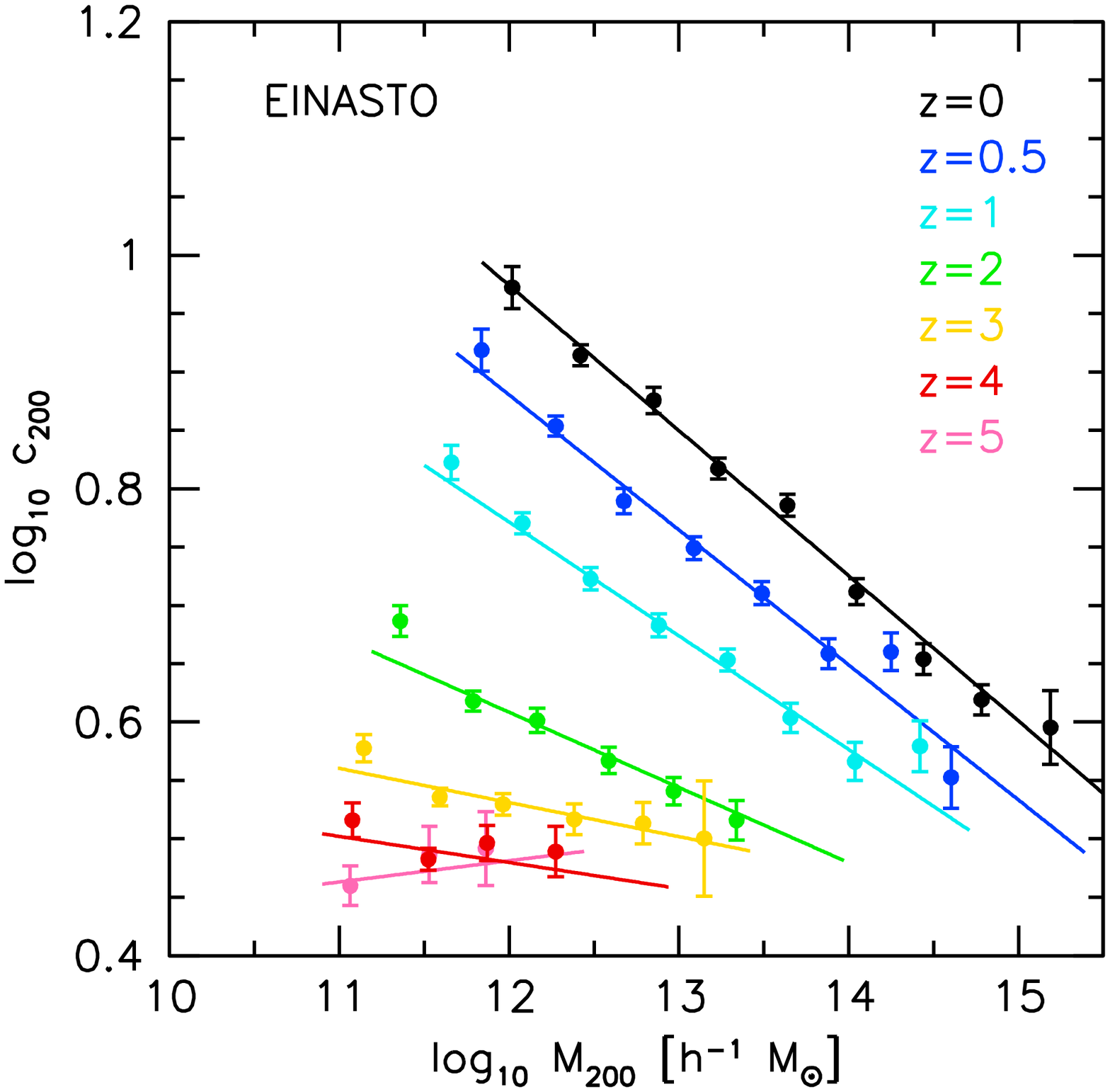,width=0.45\textwidth}
\psfig{figure=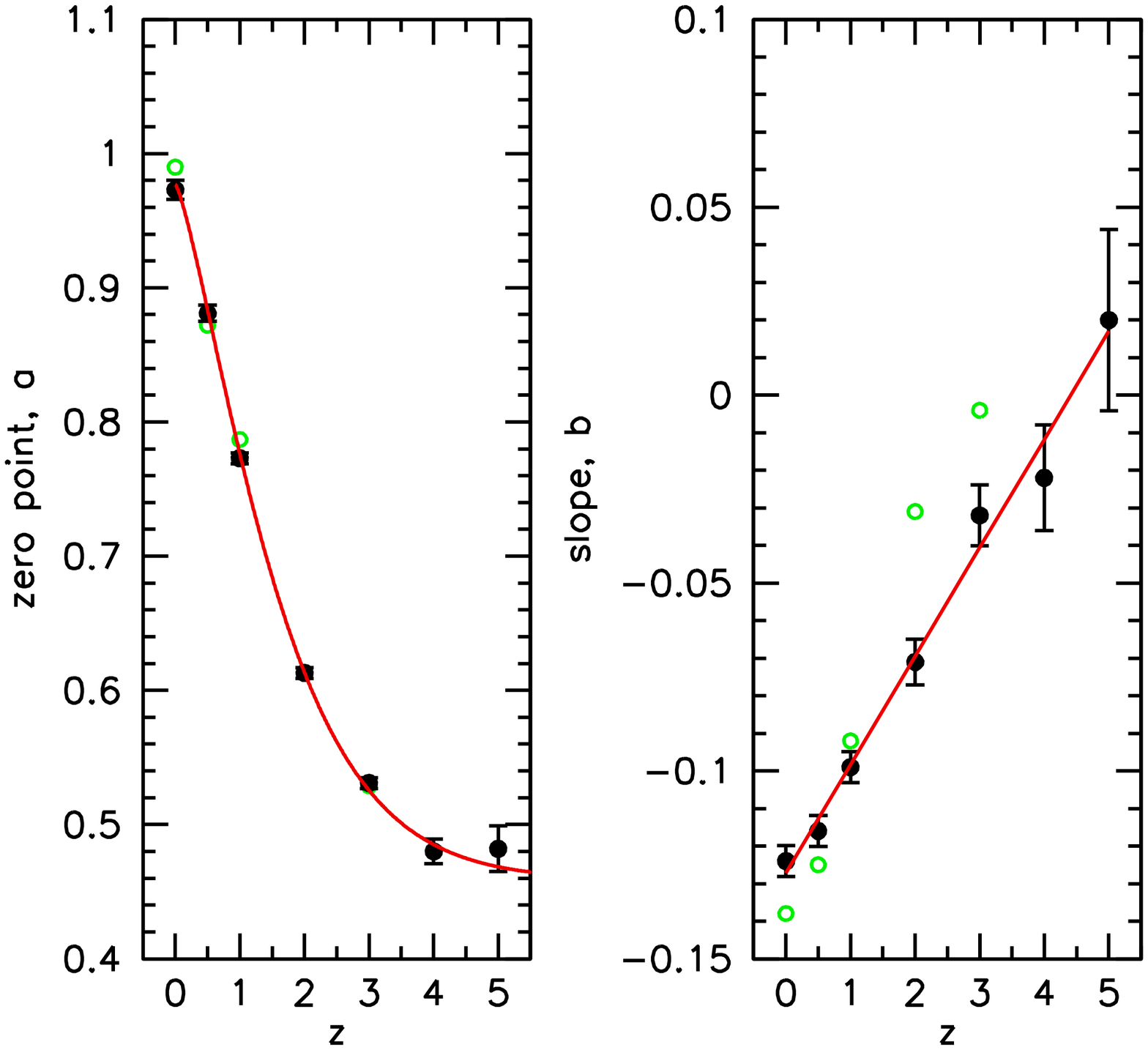,width=0.45\textwidth}
}
\caption{Evolution of the concentration mass relation using Einasto
  fits and the $c_{200}=r_{200}/r_{-2}$ definition.  The left panels
  show the median concentration in bins of halo mass from redshifts
  $z=0$ to $z=5$ (with colors as indicated). The solid lines are
  power-law fits of the form $\log_{10}c_{200} = a +
  b\log_{10}(M_{200}/10^{12}\hMsun)$. The parameters of these fits are
  given as black points in the right panels, with the error bars
  showing $1\sigma$ uncertainties. As with the NFW concentrations
  (Fig.~\ref{fig:cmz_nfw}) the evolution of the slope is linear with
  redshift, while the evolution of the zero point is approximately
  exponential with redshift. The red solid lines give fitting formula
  for the evolution (see Eqs.~\ref{eq:cm_slope_einasto} \&
  \ref{eq:cm_zero_einasto}). The green open circles show results from
  Gao \etal (2008) for the WMAP1 cosmology (for redshifts $z=0$ to
  $z=3$), who find a similar zero point evolution, but stronger slope
  evolution.}
\label{fig:cmz_ein_fit}
\end{figure*}

The upper right panel shows the Navarro, Frenk, \& White (1997) model
using the modified parameters $F=0.1$, $f=0.01$, $C=600$ from Gao
\etal (2008). This model does a better job than the original NFW model
at reproducing the evolution of concentrations for low mass haloes,
but this is at the expense of under predicting the concentrations of
high mass haloes.

The lower left panel shows the Bullock \etal (2001) model as modified
by Macci\`o \etal (2008). As in Fig.~\ref{fig:bullock} we use
$K=4.2$. The success of this model is in reproducing the slope of the
concentration mass relation for low mass haloes at all redshifts. It
also does quite well in reproducing the zero point evolution. The
model fails in reproducing the concentrations of high mass haloes. An
approximate solution to this problem is to set the concentration to
the parameter $K$ at redshifts before the collapse time of the halo.

The lower middle panel shows the Zhao \etal (2009) model. Overall,
this fares better than the other models, but not as well as our
fitting formula. Note that in the Zhao \etal (2009) model there is a
minimum concentration of $\cvir=4$. Since the ratio between
$\cvir/c_{200}$ increases towards lower redshifts, this results in
their model predicting the lowest $c_{200}$ in the highest mass haloes
at $z=0$. We have verified that their model still under predicts the
concentrations of high mass haloes at low redshifts when we plot
$\cvir$ vs $\Mvir$.

The lower right panel shows the Prada \etal (2012) model, which unlike
all of the other analytic models is empirical. This model was
calibrated against a WMAP 5th year cosmology, but it does include some
cosmology dependence, so in principle it should apply to different
cosmologies. The unique feature of his model is an upturn in halo
concentrations in high mass haloes, which is not clearly seen in our
simulations. The main success of this model is in reproducing the
slope evolution for haloes with masses $M_{200} \lta 10^{13}\Msun$,
however, it predicts weaker evolution than we find in our
simulations. 

One of the features of the model that does not depend on cosmology is
the minimum halo concentration which is $c_{200}^{\rm min}=3.681$ at
very high redshift and $c_{200}^{\rm max}\simeq 5.0$ at redshift
$z=0$. The discrepancy with our simulations suggest that the minimum
concentration may be cosmology dependent: at redshift $z\sim 4$ the
minimum concentration of their model is $c_{200}^{\rm min}\simeq 4.0$,
whereas our simulations have $c_{200}^{\rm min} \simeq
3.5$. Alternatively, at least part of the discrepancy is likely due to
systematic errors in estimating halo concentrations from $V_{\rm
  max}/V_{200}$, which we have shown gives systematically higher
concentrations than NFW or Einasto fits to the density profile.

In summary none of the analytic models in the literature accurately
reproduce the evolution of the concentration mass relation as found in
our simulations. Thus, for the purpose of predicting the
concentrations of galaxy mass haloes since redshift $z=5$, our simple
fitting formula provide the most accurate description currently
available. If the concentrations are desired for halo masses and
redshifts far outside the range of our simulations, then one of the
analytic models discussed above would likely be preferred.

\section{Evolution of Einasto parameters}
\label{sec:evolution_einasto}
In this section we discuss the evolution of structural parameters from
Einasto fits to dark matter haloes: scale radii as expressed through
the concentration parameter, $c$, and the Einasto shape parameter,
$\alpha$.  These provide a more accurate description of CDM halo
profiles than the NFW function (see \S~\ref{sec:einasto}), and thus
should be used whenever possible.  Recall, as discussed in
section~\ref{sec:resolution}, in order to recover robust parameters we
require more particles for Einasto fits than for NFW ones, and more
particles for lower redshift haloes. This results in redshift
dependent lower halo mass limits.

\subsection{Concentration mass relation}
Fig.~\ref{fig:cmz_ein_fit} shows the evolution of the concentration
mass relation obtained from Einasto fits, and using the $\Delta=200$
halo definition. The evolution of the slope and zero point is given by
\begin{equation}
\label{eq:cm_slope_einasto}
b = -0.130 +0.029z,	
\end{equation}
\begin{equation}
\label{eq:cm_zero_einasto}
a = 0.459 +(0.977-0.459) \exp(-0.490 z^{1.303}).
\end{equation}
The slope evolution is consistent with being linear in redshift, which
implies a positive slope at $z>4.5$. Our measurement at $z=5$ is
consistent with the linear evolution, but also, within the errors with
a slope of zero. Further work is needed to constrain the behaviour of
the slope evolution at very high redshift.
For comparison, the green points in the right hand panels of
Fig.~\ref{fig:cmz_ein_fit} show results from $z=3$ to $z=0$ from Gao
\etal (2008) who used the Millennium Simulation.  Their results have a
very similar zero point evolution as ours, but a stronger slope
evolution.

\subsection{Einasto shape parameter}
\begin{figure}
\centerline{
\psfig{figure=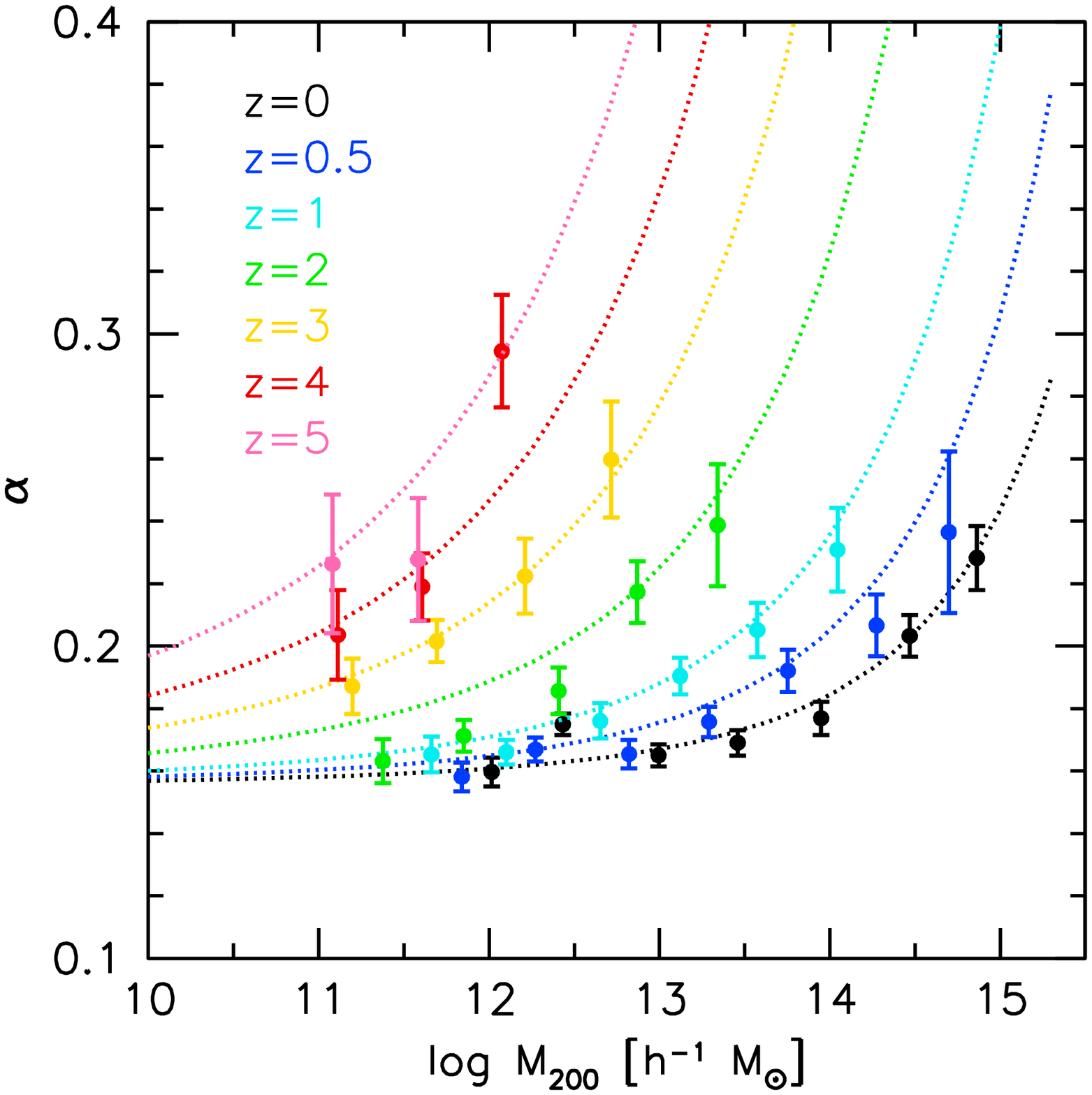,width=0.45\textwidth}
}
\centerline{
\psfig{figure=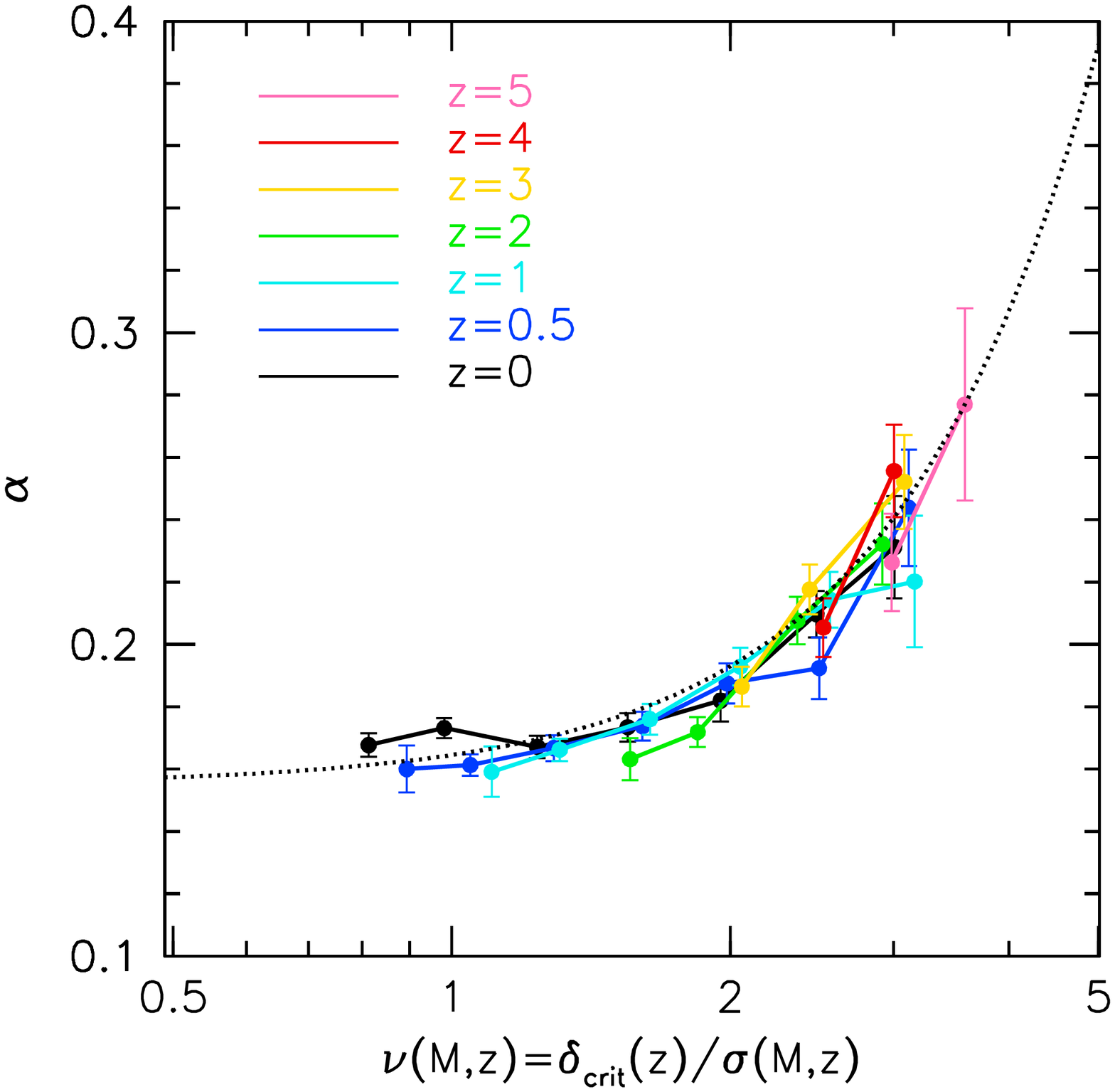,width=0.45\textwidth}
}
\caption{Evolution of median Einasto shape parameter, $\alpha$, versus
  halo mass (upper panel) and dimensionless peak height parameter
  $\nu(M,z)=\delta_{\rm crit}(z)/\sigma(M,z)$ (lower panel).  The
  dotted lines are the relation from Gao \etal (2008), which also
  provides a good fit to our simulation data in spite of the different
  cosmological parameters.}
\label{fig:amz}
\end{figure}

The upper panel of Fig.~\ref{fig:amz} shows the evolution of the
Einasto shape parameter, $\alpha$, as a function of halo mass.  At low
redshifts a Milky Way mass halo ($M_{200}\sim 10^{12}\hMsun$) has on
average $\alpha\simeq 0.16$, which is consistent with results from
much higher resolution simulations of individual haloes (Stadel \etal
2009; Navarro \etal 2010).  At higher masses $\alpha$ increases
gradually until $M_{200}\sim 10^{14}\hMsun$, above which $\alpha$
increases rapidly.  This mass dependence qualitatively explains the
differences between inner density profile slopes obtained from our NFW
and Einasto fits shown in Fig.~\ref{fig:einasto_vs_nfw}.  From
Fig.~\ref{fig:einasto_vs_nfw2} we expect haloes to be denser than NFW
at small radii, with a larger difference at lower masses -- which is
exactly what we find in Fig.~\ref{fig:einasto_vs_nfw}.

The mass dependence of the Einasto shape parameter is present at all
redshifts we probe. At fixed halo mass $\alpha$ increases with
redshift, such that at redshift $z\sim 4$ haloes of Milky Way mass
have $\alpha\simeq0.25$.  As shown by Gao \etal (2008) this evolution
effectively disappears when the halo virial mass is replaced by the
dimensionless peak height, $\nu(M,z)$:
\begin{equation}
  \nu(M,z)=\delta_{\rm crit}(z)/\sigma(M,z).
\end{equation}
Where $\delta_{\rm crit}(z)$ is the ratio of the linear density
threshold for collapse at redshift $z$, and $\sigma(M,z)$ is the rms
linear density fluctuation at $z$ within spheres of mean enclosed
mass, $M$. The parameter $\nu(M,z)$ is related to the abundance of
objects of mass $M$ at redshift $z$. The characteristic mass $M_*(z)$
of the halo mass distribution at redshift $z$ is defined through
$\nu(M_*,z)=1$.

In the lower panel in Fig.~\ref{fig:amz} we confirm this result and
extend it to higher redshifts (the highest redshift analyzed by Gao
\etal 2008 was $z=3$). The dotted line shows the relation between $\alpha$
and $\nu$ from Gao \etal (2008) and is given by
\begin{equation}
\label{eq:alpha_nu}
\alpha=0.0095\nu^2 +0.155.
\end{equation}
This provides a good match to our simulation results, even though the
cosmological parameters are significantly different.  The dotted lines
in the upper panel of Fig.~\ref{fig:amz} show the predicted $\alpha$
vs $M_{200}$ relation from Eq.~\ref{eq:alpha_nu}. Since these
relations require a numerical calculation, in Appendix~\ref{sec:numz}
we provide accurate fitting formulae for calculating the $\nu(M,z)$ in
the Planck cosmology.

\begin{figure}
\psfig{figure=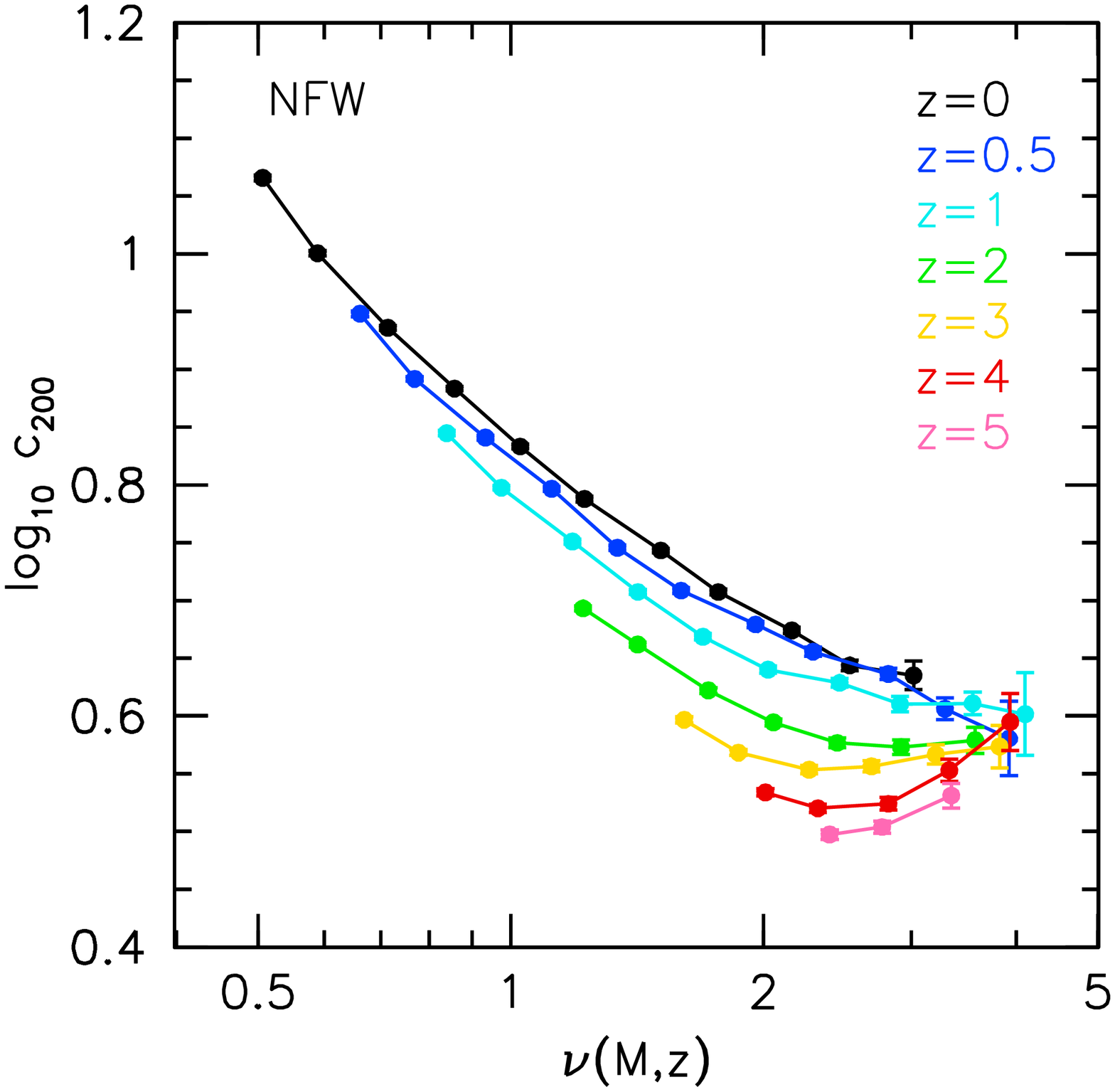,width=0.45\textwidth}
\psfig{figure=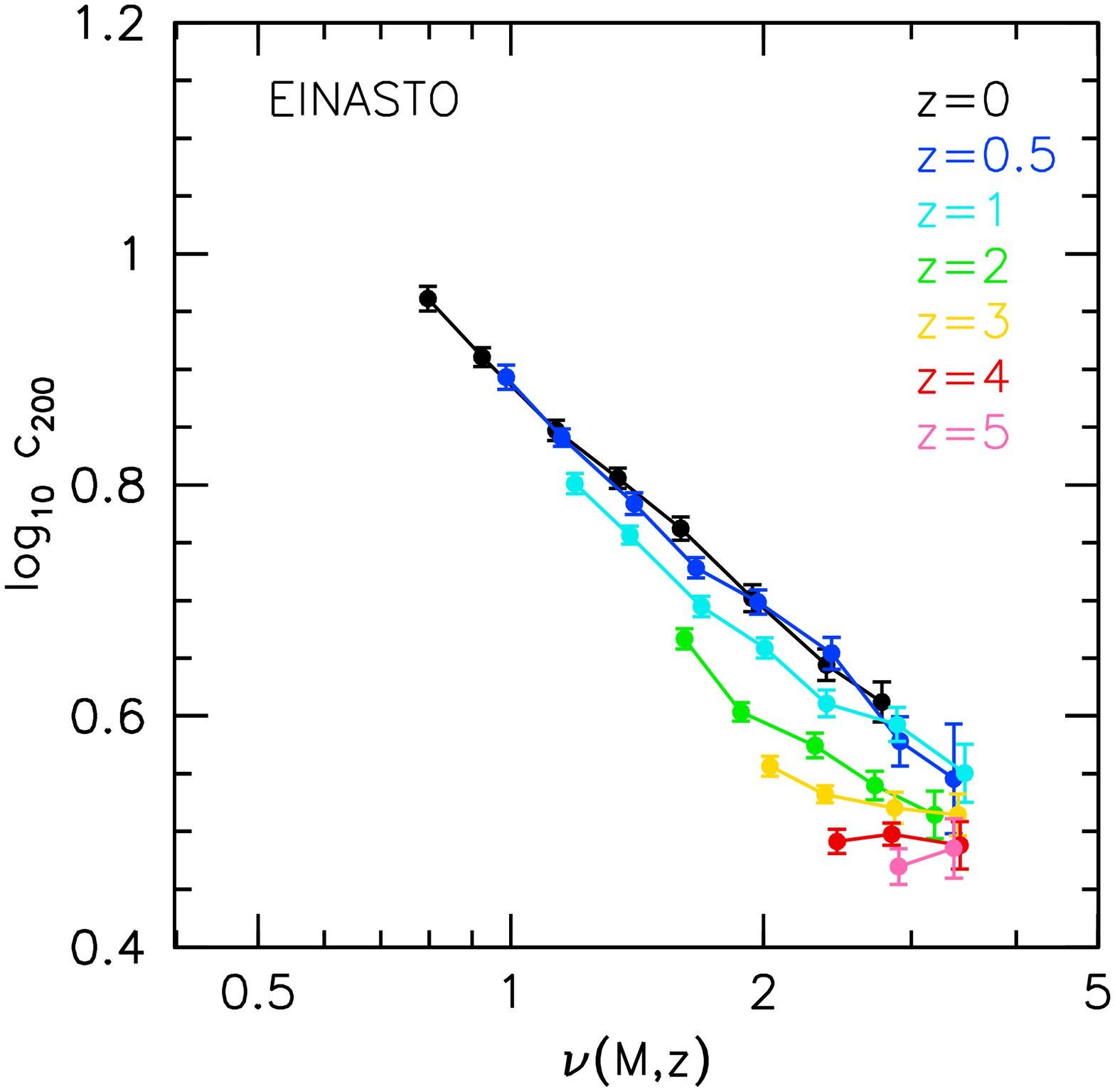,width=0.45\textwidth}
\caption{Evolution of median concentration parameter,
  $c_{200}=r_{200}/r_{-2}$ as a function of dimensionless peak height
  parameter $\nu(M,z)$ for NFW fits to haloes with more than 500
  particles (upper panels) and Einasto fits to haloes with $N_{\rm
    min}$ as describe in \S\ref{sec:resolution} (lower panels). In
  contrast to the relation between Einasto shape parameter and halo
  mass (Fig.~\ref{fig:amz}), the evolution in the concentration mass
  relation is not removed when replacing halo mass with peak height
  parameter. }
\label{fig:cnuz_nfw}
\end{figure}

\subsection{Concentration vs peak height}
The concentration vs halo mass relation evolves strongly. However, as
shown by previous authors (Prada et al. 2012; Ludlow et al. 2013b)
there is much less evolution when the halo mass is replaced by the
dimensionless peak height parameter, $\nu(M,z)$.
 
Using the set of Millennium Simulations (which assume a cosmology
similar to WMAP1) Ludlow \etal (2013b) argue that the $c_{200}-\nu$
relation has no redshift dependence (since $z=2$). They use this as a
basis for an analytic model for the evolution of halo
concentrations. Using a wider range of simulations and cosmologies,
Prada \etal (2012) find significant redshift dependence (since
$z=10$).  Fig.~\ref{fig:cnuz_nfw} shows the evolution of the
concentration vs peak height relation from redshifts $z=5$ to $z=0$
from our simulations. In agreement with Prada \etal (2012), but in
disagreement with Ludlow \etal (2013b), we find that this relation is
not redshift independent. For example, at $\nu=2$ (i.e., a $2\sigma$
mass fluctuation) we find a factor of 1.6 increase in concentration
from $z=4$ to $z=0$, which is much larger than can be attributed to
systematic errors in measuring halo concentrations. 

In light of the redshift independent relation between Einasto shape
parameter, $\alpha$, and $\nu$ it is natural to ask whether
$r_{200}/r_{-2}$ is the correct way to define halo
concentrations. Note that other commonly used definitions of the
virial radius, will only result in more evolution.  For example, at
redshift zero $\cvir/c_{200} \sim 1.3$, while at high redshift
$\cvir/c_{200}\sim 1$.

\begin{figure}
\psfig{figure=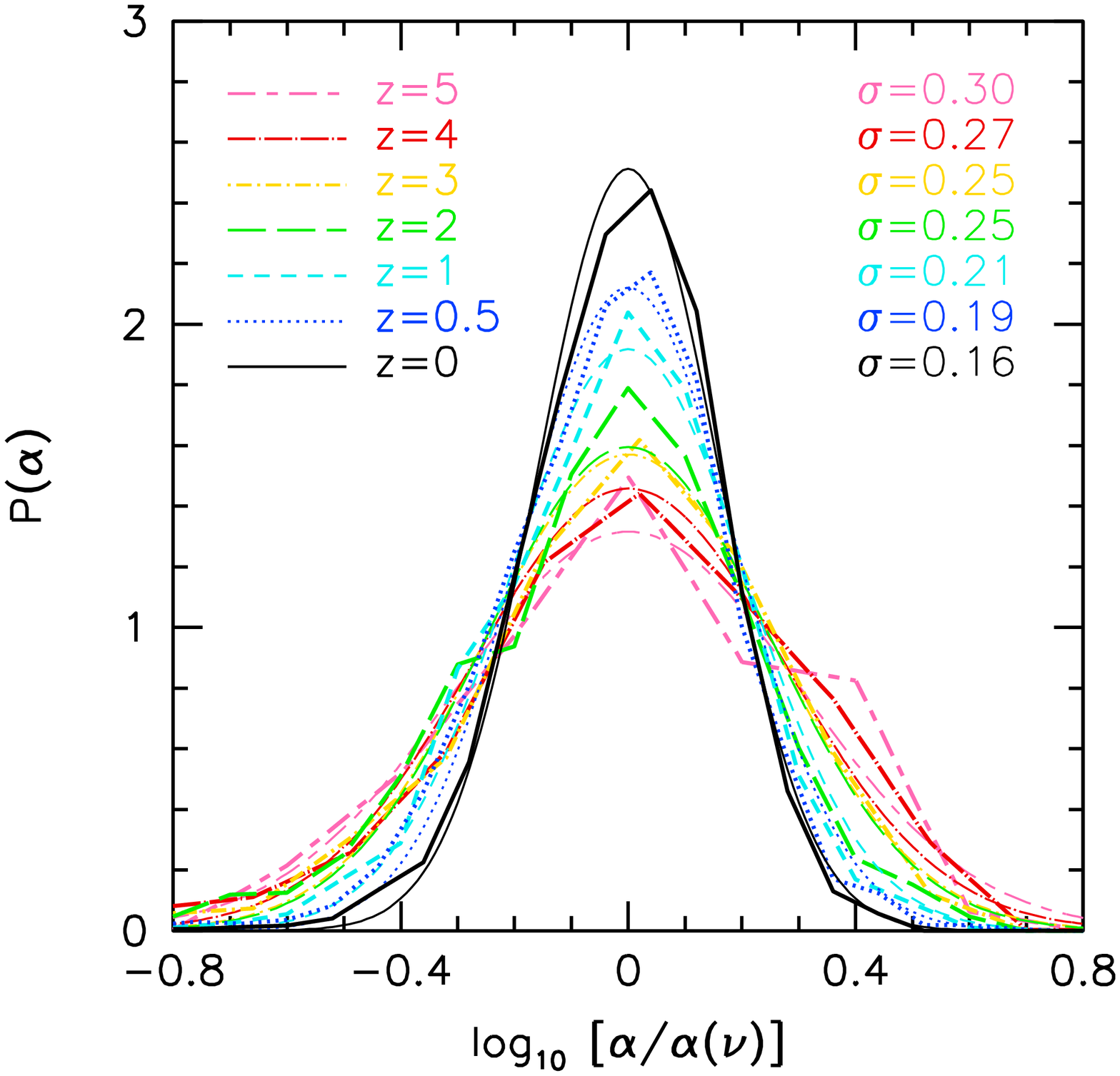,width=0.45\textwidth}
\psfig{figure=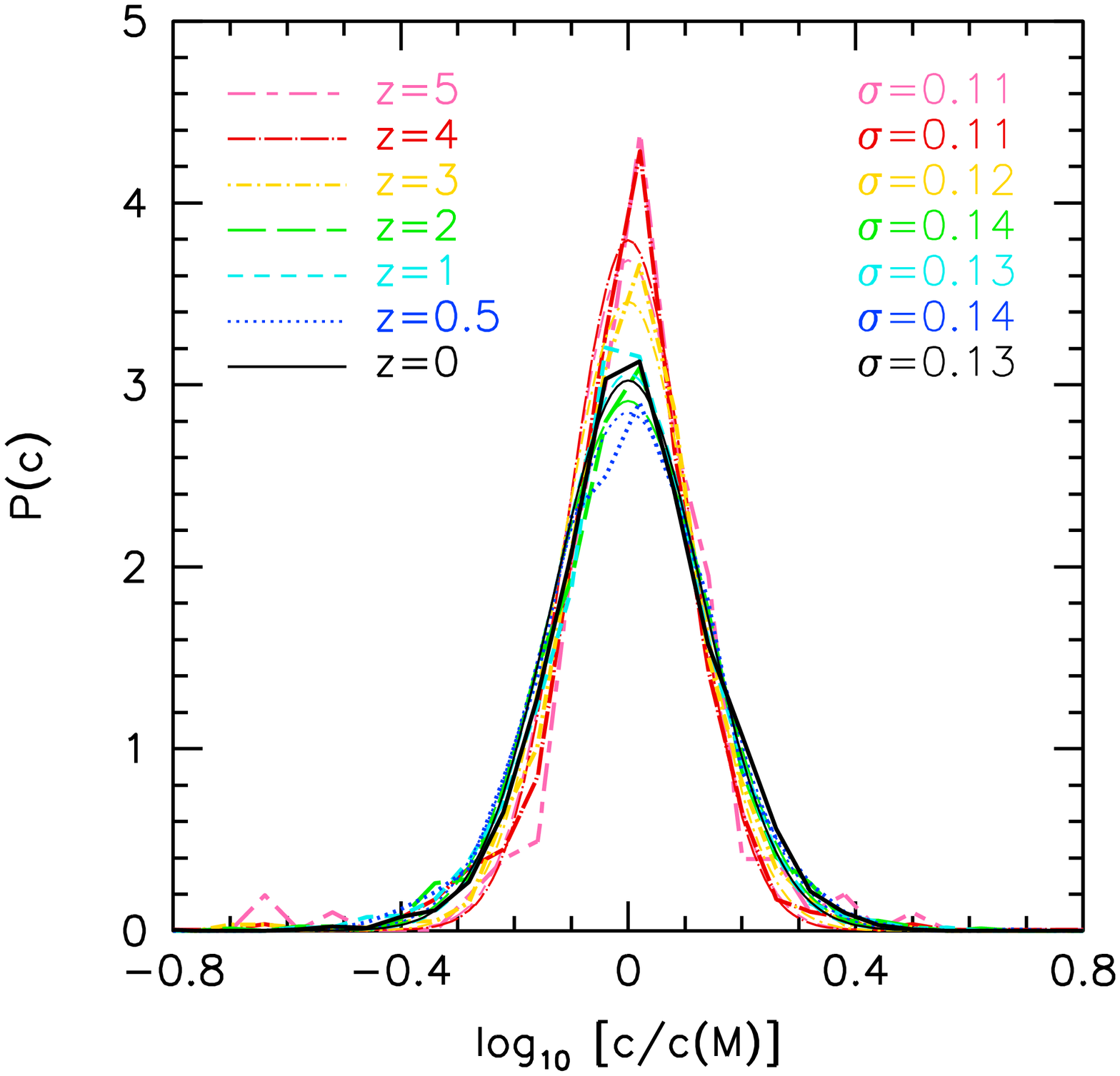,width=0.45\textwidth}
\caption{Scatter in the Einasto concentration, $c$, and shape
  parameter, $\alpha$, as a function of redshift. The distributions of
  $c$ and $\alpha$ are well described by log-normals with scatter of
  $\sim 0.13$ dex and $\sim 0.2$ dex, respectively.}
\label{fig:danu}
\end{figure}

\subsection{Scatter in structural parameters}
Gao \etal (2008) fitted Einasto profiles to stacks of haloes in bins
of halo mass. Here, we fit Einasto profiles to individual haloes, and
thus can measure the scatter in $\alpha$.  The distribution of
$\alpha(\nu)$ and $c(M)$ are shown in Fig.~\ref{fig:danu}.  Both
distributions are well described by log-normal functions.  For $c$ the
standard deviation is roughly independent of redshift at 0.13 dex. For
$\alpha$ the measured standard deviation increases with redshift
roughly following $\sigma_{\log_{10}\alpha}=0.16+0.03z$.  Thus the
halo mass and redshift, or equivalently the peak height parameter, are
not sufficient to fully specify the structure of CDM haloes.

\begin{figure*}
\centerline{
\psfig{figure=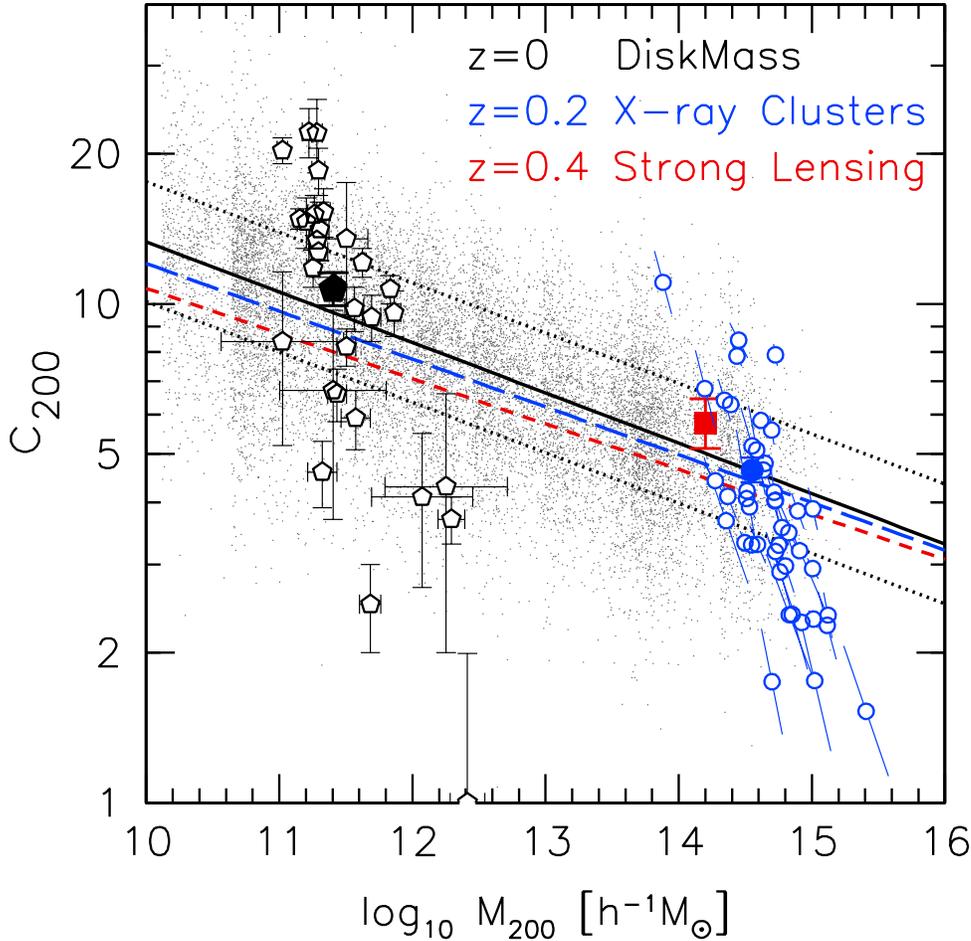,width=0.75\textwidth}
}
\caption{Comparison between the NFW concentration mass relation from
  our simulations (lines and points) with observations of spiral
  galaxies (black pentagons) and clusters of galaxies (blue circles,
  red square) at low redshifts. For each cluster there are two
  measurements. We plot the average and connect the two measurements
  with a line.  This demonstrates the fitting degeneracy between
  concentration and halo mass. For the spirals and clusters the filled
  symbols show the median concentration and halo mass. Combining the
  observational data sets yields a slope and zero point in good
  agreement with our simulations, implying only mild halo response to
  galaxy formation at these two scales.}
\label{fig:diskmass}
\end{figure*}

One must be wary of covariances between parameters when fitting
Einasto profiles. For example, Ludlow \etal (2013a) showed that at
fixed halo mass $\alpha$ and concentration are negatively correlated
-- haloes with higher fitted $\alpha$ have lower concentration. We
also find such a correlation when using Einasto fit
parameters. However when using NFW concentrations and Einasto $\alpha$
there is very little correlation.  Thus at this stage we do not wish
to over-interpret the amount of scatter in $\alpha$, nor its
correlation with the concentration parameter.


\section{Comparison with observations}
\label{sec:observations}
In Fig.~\ref{fig:diskmass} we compare the concentration mass relation
from our simulations with observational measurements, both using NFW
fits. The points show a sampling of individual haloes from our
simulations at redshift $z=0$. The solid black line shows our fitting
formula for NFW profiles, with the dotted lines showing $\pm 1\sigma$
scatter. The blue long-dashed and red short-dashed lines show the
corresponding relations for $z=0.2$ and $z=0.4$, respectively. The
black pentagons show observational results for spiral galaxies at
redshift $z\simeq 0$ from the DiskMass project (Bershady \etal 2010;
Martinsson \etal 2013). The blue circles show results for x-ray
luminous clusters at redshifts $z=0.1$ to $0.3$ from Ettori \etal
(2010). The red square shows the mean mass and concentration from an
analysis of strong lensing groups and clusters by Auger \etal (2013),
which have a mean redshift of $z\simeq 0.4$.  Recall the uncertainty
on cosmological parameters imparts just a 3.5\% uncertainty to the
concentrations of Milky Way mass haloes, which is much smaller than
the systematic measurement uncertainties in both observation and
theory.

Individually, the results from spirals and clusters imply a very steep
slope to the concentration mass relation. However, as shown by Auger
\etal (2013), the steep slope is almost entirely due to the covariance
between halo mass and concentration. This can also be seen in the
x-ray cluster data from Ettori \etal (2010), since the authors use two
methods to estimate halo masses and concentrations. In
Fig.~\ref{fig:diskmass} the two numbers for each cluster are connected
by a diagonal line, which has a mean slope of $-0.82$.
 
The median concentrations and masses for the spirals and x-ray
clusters are shown with a solid pentagon and solid circle,
respectively. Together these imply a slope to the observed
concentration mass relation of $\simeq -0.1$. In addition, since both
points are close to the median concentration-mass relations from our
simulations, this suggests that dark matter haloes have, on average,
experienced little response to galaxy formation.  While there are
several caveats in directly interpreting this plot, it shows that the
\LCDM paradigm is consistent with the observations of massive dark
matter haloes.

\section{Summary}
\label{sec:summary}

In this paper we have used a large set of cosmological N-body
simulations to study the evolution of the structure of cold dark
matter haloes over 90\% of the age of the Universe. At redshift zero
our simulations span five orders of magnitude in halo mass
($10^{10}-10^{15} \hMsun$), covering haloes those that host individual
dwarf galaxies to those associated with massive clusters. We adopt the
cosmological parameters derived from the first data release of the
Planck Satellite (the Planck Collaboration 2013).
We summarize our results as follows:
\begin{itemize}
\item The concentration mass relation in the Planck cosmology has a
  20\% higher normalization (at redshift $z=0$) than in the WMAP 5th
  year cosmology. Despite significant differences in cosmological
  parameters the Planck concentration mass relation is very similar to
  that from the WMAP1 cosmology. By coincidence, the increased
  $\Omega_{\rm m}$ (in the Planck vs WMAP1 cosmology) is almost
  perfectly balanced by the decrease in $\sigma_8, n$, and $h$.
\item Propagating the uncertainties in cosmological parameters given
  by the Planck Collaboration results in just a 3.5\% uncertainty in
  the concentrations of Milky Way mass haloes at redshift $z=0$, which
  is smaller than typical systematic uncertainties in measuring halo
  concentrations.
\item In agreement with previous studies we find that the spherically
  averaged density profiles of CDM haloes are better described by the
  Einasto (1965) profile than the NFW (1997) profile. For example,
  between 2\% and 4\% of the virial radius, simulated haloes of mass
  $M_{200}\sim10^{13}\hMsun$ (which host massive elliptical galaxies)
  have average logarithmic density slopes $d\log\rho/d\log r \simeq
  -1.6$, compared to $\simeq -1.4$ for NFW fits.
\item At fixed redshift the average Einasto shape parameter, $\alpha$,
  increases with halo mass, from $\alpha\sim 0.16$ for dwarf haloes to
  $\alpha\sim 0.25$ for cluster haloes. At fixed halo mass the
  average, $\alpha$, increases systematically with redshift. This
  evolution is well described by the relation between $\alpha$ and
  dimensionless peak height $\nu(M,z)=\delta_{\rm
    crit}(z)/\sigma(M,z)$ proposed by Gao \etal (2008) --- who used
  simulations in the WMAP1 cosmology.
\item The distribution in $\alpha$ about the $\alpha-\nu$ relation is
  well described by a log-normal function, with a standard deviation
  in $\log_{10}\alpha$ of $\sim 0.2$ and a slight dependence on
  redshift.
\item We find systematic differences of order $\sim 10\%$ in halo
  concentrations due to different fitting methods. Relative to Einasto
  fits to the density profile, NFW fits give concentrations that
  differ by up to 15\%, while the $V_{\rm max}/V_{200}$ method (e.g.,
  Klypin \etal 2011; Prada \etal 2012) gives differences of up to
  25\%. A consequence of these systematic differences is the upturn in
  the concentration mass relation at high masses and redshifts (e.g.,
  Klypin \etal 2011; Prada \etal 2012) is not present for our Einasto
  fits.
\item None of the analytic models in the literature (e.g., NFW 1997;
  Bullock \etal 2001; Gao \etal 2008; Zhao \etal 2009; Prada \etal
  2012) accurately reproduce the evolution of the concentration mass
  relation. We provide simple fitting formulae for NFW
  (Eqs.~\ref{eq:c200_slope}--\ref{eq:cvir_zero}) and Einasto fits
  (Eqs.~\ref{eq:cm_slope_einasto} \& \ref{eq:cm_zero_einasto}), which
  are valid between redshifts $z=5$ and $z=0$.
\item The observed concentrations and halo masses from NFW fits to
  data of spiral galaxies from the DiskMass project (Bershady \etal
  2010; Martinsson \etal 2013), groups and clusters of galaxies
  (Ettori \etal 2010; Auger \etal 2013) are in good agreement with our
  simulations suggesting only mild halo response to galaxy formation
  on these scales.
\end{itemize}

\section*{Acknowledgements} 
We acknowledge support from the Sonderforschungsbereich SFB 881 ``The
Milky Way System'' (subproject A1) of the German Research Foundation
(DFG).  The numerical simulations were performed on the {\it theo}
supercomputer at the Max-Planck-Institut f\"ur Astronomie at the
Rechenzentrum in Garching, and on the {\it Milky Way} supercomputer,
co-funded by SFB 881 (subproject Z2) and by the J\"ulich
Supercomputing Center (JSC). We thank Phil Marshall for illuminating
discussions about the versatility of the Einasto profile, and the
Referee for providing constructive and prompt comments on the
manuscript.


\appendix
\section{Peak height vs halo mass and redshift }
\label{sec:numz}
The relation between halo mass, $M$, and dimensionless peak height
parameter, $\nu(M,z)=\delta_{\rm crit}(z)/\sigma(M,z)$, requires a
numerical calculation. For convenience here we provide a fitting
formula for the relation between halo mass, peak height and
redshift. Since the redshift dependence of $\nu(M,z)$ is independent
of halo mass, we split the relation into two parts:$\nu(M,z)=\nu(M,0) \times 
[\nu(M,z)/\nu(M,0)]$.  

\begin{center}
\begin{figure}
\psfig{figure=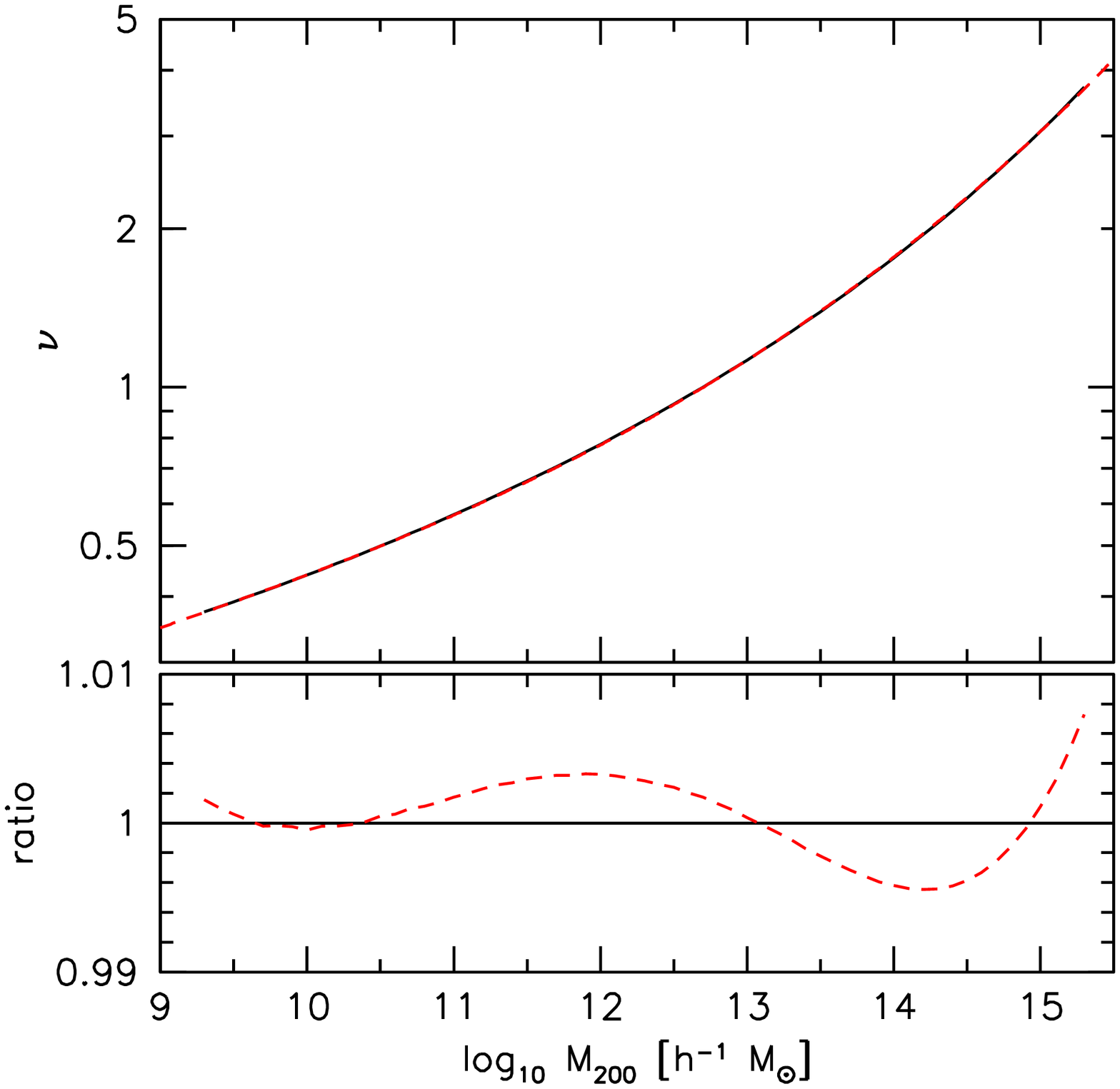,width=0.45\textwidth}
\caption{Relation between dimensionless peak height, $\nu$, and halo
  mass, $M_{200}$ (upper panel). The black solid line shows the numerical
  calculation while the red dashed line shows our fitting formula
  Eq.~\ref{eq:num}, which is accurate to better than 1\% for the mass
  range plotted (lower panel).}
\label{fig:num}
\end{figure}
\end{center}

\begin{center}
\begin{figure}
\psfig{figure=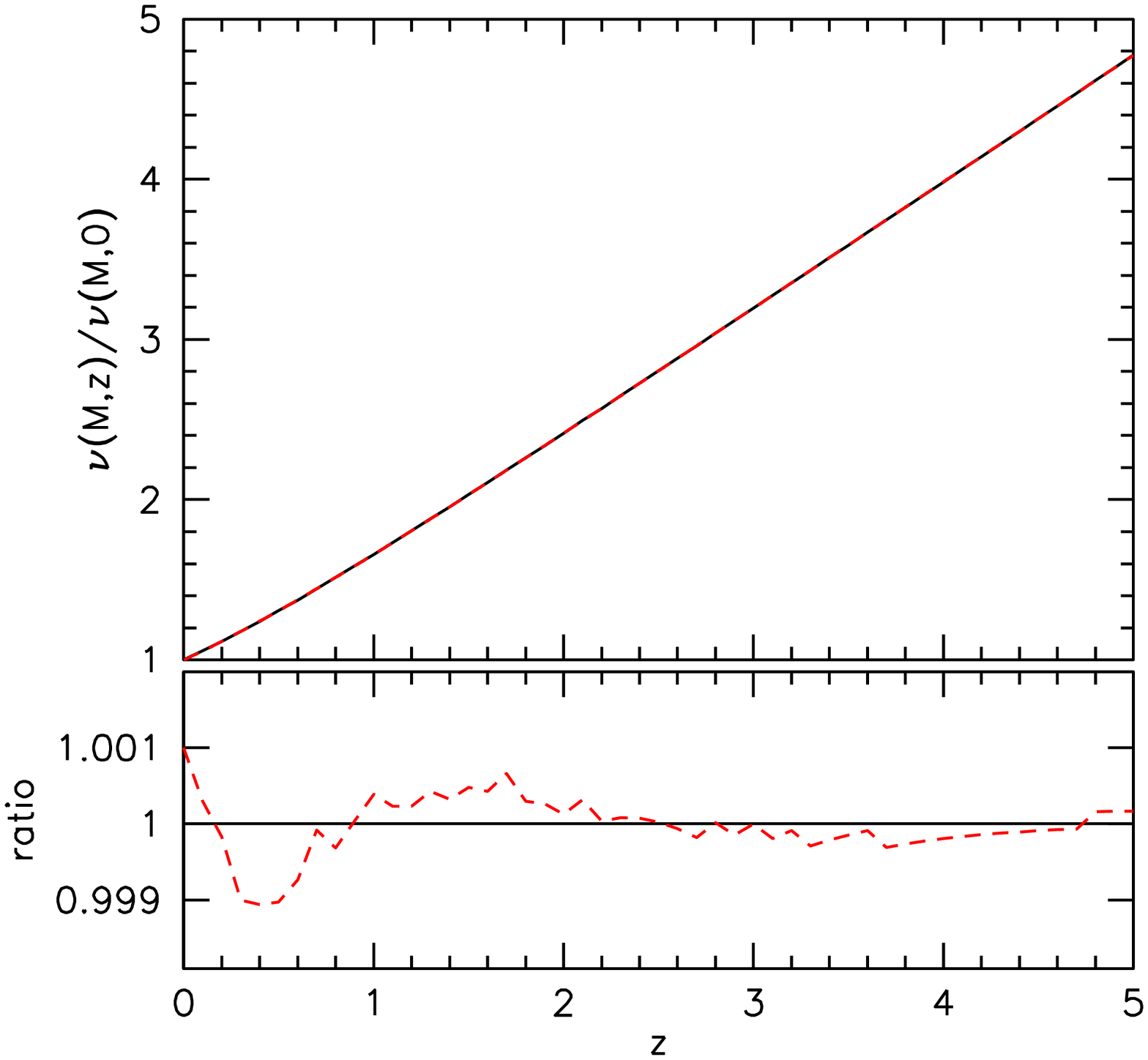,width=0.45\textwidth}
\caption{Evolution of dimensionless peak height, $\nu$ (upper panel). The black
  solid line shows the numerical calculation while the red dashed line
  shows our fitting formula Eq.~\ref{eq:nuz}, which is accurate to
  better than 0.1\% for the redshift range shown (lower panel).}
\label{fig:nuz}
\end{figure}
\end{center}

The relation between peak height and halo mass at redshift zero is
shown in Fig.~\ref{fig:num} and is approximated by
\begin{equation}
\label{eq:num}
\log_{10}\nu= -0.11 +0.146m+0.0138m^2 +0.00123m^3
\end{equation}
where $m=\log_{10}(M_{200}/10^{12}\hMsun)$. As shown in the lower
panel of Fig.~\ref{fig:num} this approximation reproduces the
numerical calculation to better than 1\% for halo masses $10^{9} \lta
M_{200}/[\hMsun] \lta 10^{15}$.  The redshift evolution of $\nu$ is
shown in Fig.~\ref{fig:nuz} and is approximated by
\begin{equation}
\label{eq:nuz}
\frac{\nu(M,z)}{\nu(M,0)}=0.033+0.79(1+z)+0.176\exp(-1.356z), 
\end{equation}
and is accurate to better than 0.1\% for redshifts $0<z<5$.  Note that
the evolution at high redshift is linear since $\OmegaM(z)\rightarrow
1$.  The deviation from a linear relation at low redshifts is due to
the increased contribution of $\OmegaL$.

Together with Eq.~\ref{eq:alpha_nu}, Eqs.~\ref{eq:num} \& \ref{eq:nuz}
can be used to calculate the redshift dependence of the average
Einasto shape parameter as a function of halo mass.

\label{lastpage}

\end{document}